\documentclass[10pt]{article}

\usepackage[utf8]{inputenc}
\usepackage[T1]{fontenc}
\usepackage[margin=0.78in]{geometry}

\usepackage{amsmath}
\usepackage{amssymb}
\usepackage{amsfonts}
\usepackage{bm}
\usepackage{booktabs}
\usepackage{longtable}
\usepackage{graphicx}
\usepackage{xcolor}
\usepackage{microtype}
\usepackage{url}
\usepackage{enumitem}
\usepackage{mathtools}
\usepackage{array}
\usepackage{hyperref}
\usepackage{float}
\usepackage{placeins}
\usepackage{flafter}
\usepackage{pdflscape}

\hypersetup{
    colorlinks=true,
    citecolor=blue,
    linkcolor=blue,
    urlcolor=blue
}

\title{
\textbf{Beyond Tokens: Probing Higher-Order Epistasis\\
in Learned Protein Representations}
}

\author{
Maryam Rahimimovassagh \\
University of Central Florida
\and
Ivan Garibay \\
University of Central Florida
\and
Niloofar Yousefi \\
University of Central Florida
}

\date{}

\begin{document}

\maketitle

\begin{abstract}

Protein fitness landscapes contain nonlinear interactions in which the
effect of one mutation depends on the identities of other residues.
Although learning systems can predict protein fitness, predictive
performance alone does not reveal how experimentally measured
higher-order structure is organized within their internal
representations.

We introduce \textbf{ORBIT}, an \textbf{Order-Resolved Benchmarking of
Interaction Transformations} framework that distinguishes interaction
presence, representation accessibility, and functional recovery.
ORBIT first validates its Walsh-based diagnostics on synthetic
landscapes with known interaction order and then analyzes the
experimentally measured GB1 fitness landscape under the biologically
motivated FLIP 2-vs-rest generalization setting. We compare ridge
regression, a standard MLP, independent token representations,
nonlinear independent tokens, and Residual Interaction Tokenization
(RIT).

In the primary two-hidden-layer comparison across 20 stochastic training
seeds, planned seed-paired inference found no architecture differences
in FLIP test $R^2$, strict third- or fourth-order functional recovery,
or final-hidden-layer third- or fourth-order accessibility. In contrast,
RIT significantly increased pairwise accessibility directly at the token
stage relative to both independent-token controls
($\Delta A_{\mathrm{tok},2}=0.2468$, $d_z=1.67$,
Holm-adjusted $p=1.14\times10^{-5}$), without a detectable downstream
higher-order advantage.

A pre-specified depth/capacity sensitivity analysis then increased the
backbone from two to three or four hidden layers. Deeper MLPs showed
significant gains in FLIP prediction, third-order functional recovery,
and final-layer third-order accessibility. Fourth-order accessibility
also improved relative to the shallow MLP, although its absolute
held-out $R^2$ remained below zero. Independent tokens showed a smaller
prediction gain at depth four, whereas nonlinear tokens and RIT showed
no Holm-significant depth effects on the confirmatory endpoints.
Because the deeper models also contain additional parameters, these
results are interpreted as depth/capacity sensitivity rather than a
parameter-matched causal effect of depth.

ORBIT therefore reveals representation-level changes that are not
apparent from conventional prediction metrics alone and separates early
interaction-aware tokenization effects from interaction structure that
can be constructed by downstream nonlinear capacity.

\end{abstract}

\section{Introduction}

Protein sequence--function relationships are often strongly nonlinear.
The phenotypic effect of an amino-acid substitution can depend on the
identities of other residues, producing \emph{epistasis}. As additional
mutations are combined, these dependencies can involve increasingly
higher-order interactions that cannot be explained by independent
single-mutation effects.

The GB1 protein fitness landscape provides a particularly useful setting
for studying this phenomenon. The original experiment measured 149,361
of the $20^4=160,000$ possible combinations across four epistatic
positions in the B1 domain of protein G
\cite{wu2016}. The landscape has subsequently become a standard
benchmark for protein fitness inference.

Machine-learning benchmarks such as FLIP explicitly use GB1 to test
whether models trained on variants with fewer mutations can generalize
to variants containing more simultaneous mutations
\cite{dallago2021}. At the same time, aggregate predictive performance
provides an incomplete description of what a model has learned.
Models with similar test error may construct, expose, and preserve
interaction structure differently inside their representations.

A complementary line of work studies fitness landscapes through their
spectral structure. Neural networks trained by gradient descent can
preferentially learn lower-degree Walsh--Fourier components
\cite{gorji2023}. Sparse spectral regularization can improve recovery of
higher-order fitness functions \cite{aghazadeh2021}, and Fourier
methods have been used to extract higher-order interactions from
protein language model outputs \cite{tsui2024}. These approaches
demonstrate that interaction order is informative at the level of the
learned function.

However, they leave a distinct representation-learning question:

\begin{quote}
\textit{Where in a molecular learning pipeline does interaction
information of increasing order become accessible, and how does that
trajectory depend on representation design and downstream capacity?}
\end{quote}

Consider a predictor composed of multiple representation stages,

\begin{equation}
X
\longrightarrow
Z_{\mathrm{tok}}
\longrightarrow
H_1
\longrightarrow
H_2
\longrightarrow
\hat{y}.
\label{eq:pipeline}
\end{equation}

Two models may achieve similar final predictive performance while
organizing interaction information very differently. An
interaction-aware tokenizer may expose pairwise relationships directly
at $Z_{\mathrm{tok}}$, whereas an initially independent representation
may rely on later nonlinear transformations to construct comparable or
higher-order structure. Likewise, increasing downstream depth or
capacity may change how much interaction structure becomes accessible
without changing the initial token representation.

We introduce \textbf{ORBIT}, an \textbf{Order-Resolved Benchmarking of
Interaction Transformations} framework for analyzing these trajectories.
ORBIT distinguishes three complementary levels:

\begin{enumerate}[leftmargin=*,itemsep=1pt]

    \item \textbf{Interaction presence:}
    which interaction orders are expressed in a learned representation;

    \item \textbf{Representation accessibility:}
    which experimentally measured interaction components can be
    recovered from each internal representation stage;

    \item \textbf{Functional recovery:}
    which interaction orders are faithfully reproduced by the final
    predicted fitness landscape.

\end{enumerate}

Presence is quantified using normalized representation Walsh energy
$E_{\ell,k}$, accessibility using held-out probe performance
$A_{\ell,k}$, and final functional recovery using order-specific
$R_k^2$. Together, these measurements allow us to distinguish a model
that merely contains order-specific variation from one in which that
information is linearly accessible or successfully expressed in the
predicted biological function.

Our central questions are:

\begin{enumerate}[leftmargin=*,itemsep=1pt]

    \item Does ORBIT correctly recover interaction order when the
    ground-truth structure is known?

    \item Can models trained only on wild-type, single, and double
    mutants recover function on experimentally measured triple and
    quadruple mutants?

    \item Where do first- through fourth-order interactions become
    accessible across learned representation stages?

    \item Does explicit interaction-aware tokenization alter the point
    at which interaction information first becomes accessible?

    \item Can predictive performance, representation accessibility,
    and higher-order functional recovery diverge?

    \item Are these conclusions sensitive to additional downstream
    network depth and capacity?

\end{enumerate}

\paragraph{Contributions.}
Our contributions are:

\begin{itemize}[leftmargin=*,itemsep=1pt]

    \item We introduce \textbf{ORBIT}, an order-resolved framework that
    jointly measures interaction presence, representation
    accessibility, and functional recovery across learned molecular
    representations.

    \item We derive a \textbf{representation-level Walsh analysis} that
    quantifies order-specific structure directly within hidden
    representations rather than only at final model outputs.

    \item We validate ORBIT on controlled synthetic landscapes with
    known first- through fourth-order structure and then apply it to the
    experimentally measured \textbf{GB1} fitness landscape under the
    biologically motivated FLIP 2-vs-rest generalization setting.

    \item We introduce an \textbf{AA-identity-strict probe protocol}
    that evaluates representation accessibility across held-out
    non-WT amino-acid identities, and we quantify stochastic variation
    across 20 paired training seeds.

    \item In the primary matched-backbone comparison, we show that
    \textbf{Residual Interaction Tokenization (RIT)} substantially
    increases pairwise accessibility directly at the token stage,
    despite no detectable downstream advantage in higher-order
    functional recovery, final-layer higher-order accessibility, or
    FLIP test $R^2$.

    \item Through a \textbf{pre-specified depth/capacity sensitivity
    analysis}, we show that additional downstream capacity can
    substantially improve third-order functional recovery and
    higher-order accessibility for the MLP. This demonstrates that
    early interaction-aware representation and later nonlinear
    construction constitute distinct routes by which interaction
    information can become accessible.

\end{itemize}

\begin{figure}[!t]
    \centering
    \includegraphics[
        width=1.5\textwidth,
        height=0.75\textheight,
        keepaspectratio
    ]{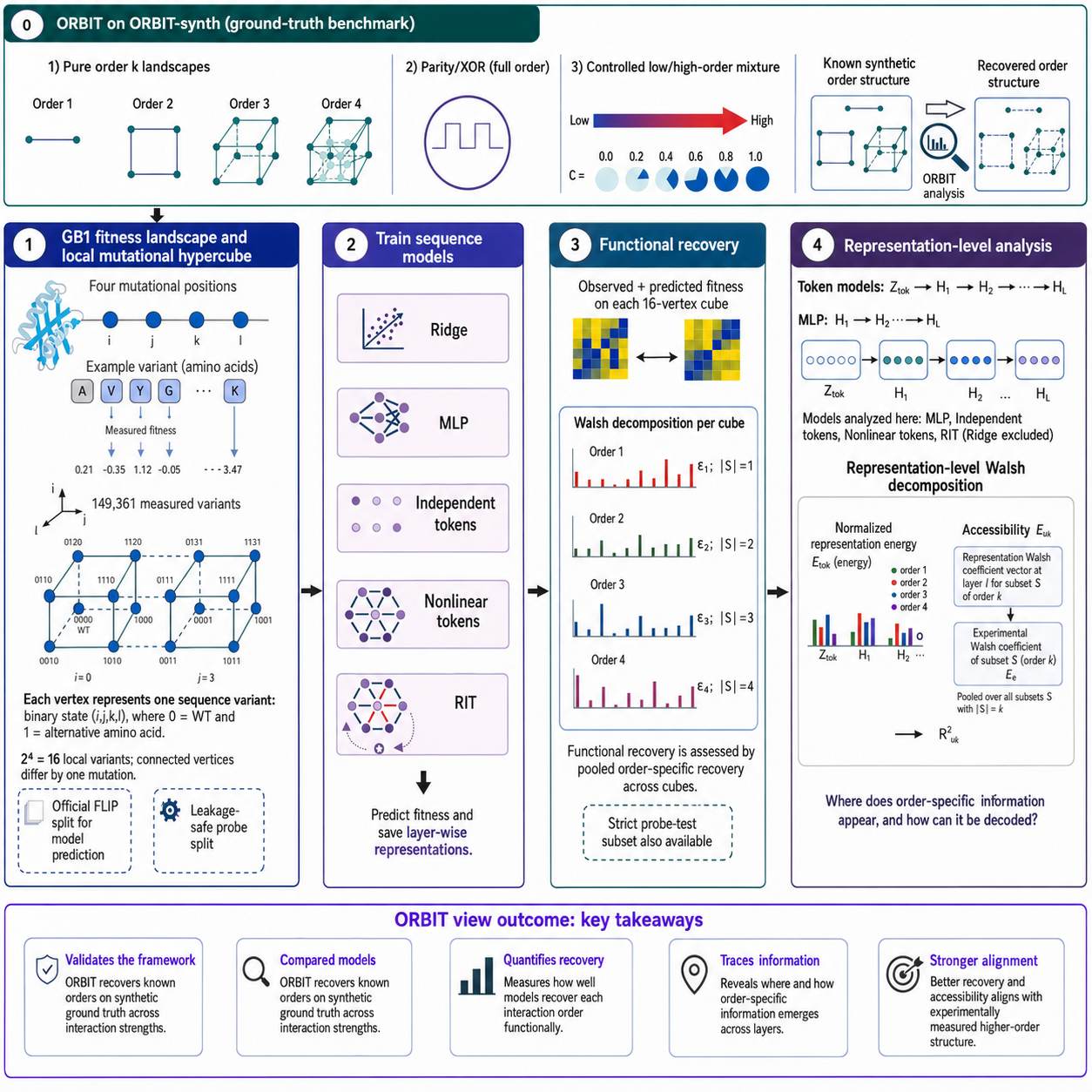}
    \caption{\textbf{ORBIT workflow.}
    ORBIT validates order-resolved diagnostics on ORBIT-Synth and then
    applies them to GB1 to compare functional Walsh recovery,
    representation energy, and linear accessibility across five models.
    The AA-identity-strict probe protocol uses disjoint non-WT amino-acid
    identities across probe partitions, and a separate pre-specified
    depth/capacity analysis extends the neural backbones.}
    \label{fig:orbit_overview}
\end{figure}

Figure~\ref{fig:orbit_overview} summarizes the primary ORBIT workflow,
from synthetic ground-truth validation to GB1 functional and
representation-level analysis. The pre-specified depth/capacity
sensitivity analysis extends the downstream backbone separately while
preserving the same ORBIT measurements.

\section{Related Work}

\paragraph{Protein fitness benchmarks.}
FLIP established standardized protein-fitness tasks for evaluating
generalization under biologically motivated data regimes
\cite{dallago2021}. For GB1, the 2-vs-rest setting trains on wild type,
single mutants, and double mutants while reserving triple and
quadruple mutants for testing. This setting motivates our primary
generalization experiment: whether models trained only on
lower-mutation-complexity variants can recover structure in unseen
higher-mutation combinations.

\paragraph{Fitness-landscape characterization.}
Predictive accuracy alone provides only a partial description of a
learned fitness landscape. GraphFLA complements prediction benchmarks
with landscape-level properties such as epistasis, ruggedness,
navigability, and neutrality \cite{huang2025}. ORBIT addresses a
different but related question: rather than characterizing only the
external predicted landscape, it tracks order-specific interaction
structure through the model's internal representations.

\paragraph{Spectral analysis of epistasis.}
Walsh--Fourier decompositions provide a natural basis for describing
interaction order in discrete biological landscapes. Neural networks
can exhibit a low-degree spectral bias \cite{gorji2023}, while sparse
spectral regularization can improve recovery of higher-order fitness
functions \cite{aghazadeh2021}. These approaches establish the value of
spectral analysis at the level of the learned function. ORBIT extends
this perspective to intermediate representations by asking where
order-specific information is present and where experimentally measured
interaction structure becomes linearly accessible.

\paragraph{Higher-order interactions in learned protein models.}
Fourier analysis has also been applied to protein language model
predictions to recover higher-order mutational interactions
\cite{tsui2024}. Such analyses examine interaction structure expressed
by model outputs. ORBIT instead follows the representation trajectory
itself, distinguishing token-stage and hidden-layer accessibility from
final functional recovery.

\paragraph{Spectral geometry of protein landscapes.}
Protein fitness landscapes can additionally be formulated over
generalized Hamming graphs, providing spectral descriptions of
ruggedness, epistasis, and learnability \cite{zhu2025}. ORBIT uses the
same underlying discrete interaction structure as an object to be
traced through learned representation space rather than only
characterized at the landscape level.

\paragraph{Higher-order structure across molecular domains.}
Beyond protein fitness landscapes, non-pairwise structure has also
motivated recent work in gene regulatory modeling. Cooperative
regulation can depend on complete sets of interacting regulators rather
than independent regulator--target effects, motivating formulations that
explicitly model non-additive set-level structure
\cite{rahimimovassagh2026beyond,rahimimovassagh2026bridge}.
These studies show that strong pairwise predictions need not imply
recovery of the underlying cooperative mechanism and that higher-order
structure can require distinct modeling and evaluation objectives.
ORBIT addresses a complementary representation-learning question:
rather than recovering interacting regulator sets, it traces where
interaction structure of increasing order is present, linearly
accessible, and ultimately expressed in the predicted protein fitness
landscape.

\paragraph{Position of ORBIT.}
Taken together, prior work provides tools for benchmarking fitness
prediction, characterizing landscape structure, and extracting
higher-order interactions from learned functions. ORBIT complements
these approaches by separating three questions that are otherwise easy
to conflate: whether order-specific structure is present in a
representation, whether it is linearly accessible, and whether it is
ultimately recovered in the predicted fitness function.

\section{ORBIT: Order-Resolved Representation Analysis}

Let a phenotype be described by

\begin{equation}
y=f(x_1,\ldots,x_d),
\end{equation}

where $x_i$ denotes the state of molecular component $i$.

For binary encodings, the landscape admits a Walsh expansion

\begin{equation}
f(x)
=
\sum_{S\subseteq[d]}
\epsilon_S \chi_S(x),
\label{eq:walsh_expansion}
\end{equation}

where

\begin{equation}
\chi_S(x)=\prod_{i\in S}x_i,
\qquad
x_i\in\{-1,+1\}.
\end{equation}

The interaction order is defined by the subset cardinality

\begin{equation}
|S|=k.
\end{equation}

For the four variable GB1 positions, the decomposition therefore
contains first-, second-, third-, and fourth-order components.

\subsection{Functional Interaction Recovery}

For a complete binary fitness cube, the Walsh coefficient associated
with subset $S$ is

\begin{equation}
\epsilon_S
=
2^{-d}
\sum_{x\in\{-1,+1\}^{d}}
f(x)\chi_S(x).
\label{eq:true_walsh}
\end{equation}

Applying the same transform to model predictions gives

\begin{equation}
\hat{\epsilon}_S
=
2^{-d}
\sum_x
\hat{f}(x)\chi_S(x).
\label{eq:pred_walsh}
\end{equation}

For each interaction order $k$, functional recovery is quantified by

\begin{equation}
R_k^2
=
1-
\frac{
\sum_{S:|S|=k}
(\hat{\epsilon}_S-\epsilon_S)^2
}{
\sum_{S:|S|=k}
(\epsilon_S-\bar{\epsilon}_k)^2
}.
\label{eq:order_r2}
\end{equation}

The resulting order-resolved functional profile is

\begin{equation}
\mathbf{R}_{\mathrm{ORBIT}}
=
(R_1^2,R_2^2,R_3^2,R_4^2).
\end{equation}

Thus, $R_k^2$ measures how faithfully the predicted fitness landscape
recovers the experimentally measured Walsh coefficients of order $k$.

\subsection{Representation-Level Walsh Components}

ORBIT extends the same decomposition to intermediate learned
representations.

Let

\begin{equation}
H_\ell(x)\in\mathbb{R}^{m_\ell}
\end{equation}

denote the representation of sequence $x$ at stage $\ell$. For each
subset $S$, define the vector-valued representation Walsh coefficient

\begin{equation}
\mathbf{r}_{\ell,S}
=
2^{-d}
\sum_x
H_\ell(x)\chi_S(x).
\label{eq:representation_walsh}
\end{equation}

where

\begin{equation}
\mathbf{r}_{\ell,S}\in\mathbb{R}^{m_\ell}.
\end{equation}

This coefficient describes representation variation associated with
the Walsh pattern indexed by $S$; it is not itself interpreted as a
biochemical interaction strength.

For interaction order $k$, the unnormalized representation Walsh energy
is

\begin{equation}
\mathcal{E}_{\ell,k}
=
\sum_{S:|S|=k}
\|\mathbf{r}_{\ell,S}\|_2^2.
\label{eq:rep_energy}
\end{equation}

We define normalized representation Walsh energy as

\begin{equation}
E_{\ell,k}
=
\frac{\mathcal{E}_{\ell,k}}
{\sum_{j=1}^{d}\mathcal{E}_{\ell,j}},
\label{eq:normalized_energy}
\end{equation}

where the order-0 constant component is excluded from the denominator.
Therefore, $E_{\ell,k}$ measures the fraction of nonconstant
representation Walsh energy associated with order $k$.

We interpret $E_{\ell,k}$ as evidence of
\emph{order-specific representation presence}, not as evidence that the
corresponding experimental epistatic coefficients are recoverable.

\subsection{Order Accessibility}

Representation energy alone does not establish whether order-specific
variation is aligned with experimentally measured epistatic structure.
We therefore separately measure \emph{accessibility}.

For each order-$k$ coefficient, a lightweight probe receives the
representation component

\begin{equation}
\mathbf{r}_{\ell,S}
\end{equation}

and predicts the corresponding experimental Walsh coefficient

\begin{equation}
\epsilon_S.
\end{equation}

The probe is

\begin{equation}
g_{\ell,k}:
\mathbf{r}_{\ell,S}
\rightarrow
\epsilon_S.
\end{equation}

Held-out probe performance defines

\begin{equation}
A_{\ell,k}
=
R^2
\left(
\epsilon_S,
g_{\ell,k}(\mathbf{r}_{\ell,S})
\right).
\label{eq:accessibility}
\end{equation}

Thus, $A_{\ell,k}$ measures how linearly accessible experimentally
measured order-$k$ structure is from representation stage $\ell$.

For the primary token models, the accessibility profile is summarized
as

\begin{equation}
\mathbf{A}
=
\begin{bmatrix}
A_{\mathrm{tok},1} &
A_{\mathrm{tok},2} &
A_{\mathrm{tok},3} &
A_{\mathrm{tok},4}
\\
A_{h_1,1} &
A_{h_1,2} &
A_{h_1,3} &
A_{h_1,4}
\\
A_{h_2,1} &
A_{h_2,2} &
A_{h_2,3} &
A_{h_2,4}
\end{bmatrix}.
\label{eq:access_matrix}
\end{equation}

The MLP has no explicit token stage and is therefore evaluated at
$H_1$ and $H_2$ only.

\paragraph{AA-identity-strict probe protocol.}
To evaluate accessibility across held-out amino-acid identities, we
partition the 19 non-WT amino-acid identities available at each of the
four GB1 positions into 10 probe-training identities, 4 validation
identities, and 5 test identities using a fixed random seed.

A WT-anchored cube is assigned to a strict partition only when all four
of its non-WT alternatives belong to that partition at their respective
positions. This yields 7,352 probe-training cubes, 245 validation cubes,
and 613 probe-test cubes. Mixed-identity cubes are excluded from strict
probe evaluation.

The non-WT amino-acid identities assigned to probe training,
validation, and test are disjoint at each position, while the WT
identity is intentionally shared. This is an
\emph{AA-identity-strict probe split}, not an upstream model-training
split: vertices belonging to probe-test cubes may have appeared in the
official FLIP model-training partition.

For each layer and interaction order, ridge-probe regularization is
selected using probe validation only. The probe is then refit on the
combined probe-training and validation partitions and evaluated once on
the 613 strict probe-test cubes. The reported $A_{\ell,k}$ is the
pooled held-out $R^2$ across experimental coefficients of interaction
order $k$.

The strict protocol therefore evaluates generalization across unseen
non-WT amino-acid identities at each position while making no claim
that the underlying protein sequences were unseen during upstream model
training.
\section{ORBIT-Synth}

We first validate the ORBIT diagnostics on synthetic binary landscapes
whose interaction order is known exactly. For $d$ binary variables,

\begin{equation}
x_i \in \{-1,+1\},
\qquad i=1,\ldots,d,
\end{equation}

we represent the landscape in the Walsh basis as

\begin{equation}
f(x)
=
\sum_{S\subseteq[d]}
\epsilon_S
\chi_S(x),
\qquad
\chi_S(x)
=
\prod_{i\in S}x_i,
\label{eq:synth}
\end{equation}

where the interaction order of coefficient $\epsilon_S$ is $|S|$.

\paragraph{Pure-order controls.}

For each interaction order $k\in\{1,2,3,4\}$, we construct a landscape
in which only Walsh coefficients satisfying $|S|=k$ are nonzero.
The nonzero coefficients are randomly sampled and normalized before
the inverse Walsh transform is applied.

Because all nonconstant energy is placed at a single order, the
normalized Walsh spectrum provides an exact ground-truth test of
whether ORBIT identifies the planted interaction order.

\paragraph{Parity control.}

We additionally construct an exact fourth-order parity landscape by
assigning a nonzero coefficient only to the full subset $S=[d]$:

\begin{equation}
f_{\mathrm{parity}}(x)
=
\prod_{i=1}^{d}x_i.
\end{equation}

Under the $\{-1,+1\}$ coding, this provides an exact full-order
parity/XOR-type interaction.

\paragraph{Controlled cooperativity sweep.}

To vary the relative contribution of lower- and higher-order structure,
we construct two unit-energy Walsh coefficient vectors,
$\boldsymbol{\epsilon}_{\mathrm{low}}$ and
$\boldsymbol{\epsilon}_{\mathrm{high}}$, occupying disjoint
interaction orders.

We combine them as

\begin{equation}
\boldsymbol{\epsilon}(C)
=
\sqrt{1-C}\,
\boldsymbol{\epsilon}_{\mathrm{low}}
+
\sqrt{C}\,
\boldsymbol{\epsilon}_{\mathrm{high}},
\qquad
C\in[0,1].
\label{eq:cooperativity}
\end{equation}

Because the two components occupy orthogonal Walsh orders, their
normalized nonconstant energy fractions are exactly

\begin{equation}
E_{\mathrm{low}} = 1-C,
\qquad
E_{\mathrm{high}} = C.
\end{equation}

For the standard $d=4$ benchmark, the lower-order component occupies
order 1 and the higher-order component occupies order 4. We evaluate

\begin{equation}
C\in
\{0.0,0.2,0.4,0.6,0.8,1.0\}.
\end{equation}

The same sampled lower- and higher-order coefficient directions are
reused across all values of $C$. Thus, the sweep changes only their
relative contribution rather than introducing a new random landscape
at each condition.

\paragraph{Role of ORBIT-Synth.}

The primary ORBIT-Synth benchmark is noiseless and contains pure
orders 1--4, exact fourth-order parity, and the six-condition
cooperativity sweep.

Its purpose is to verify that the Walsh-based ORBIT diagnostics recover
known interaction order under controlled ground truth before the same
measurements are applied to experimentally observed GB1 fitness data.
\section{GB1 Protein Fitness Landscape}

Our primary molecular experiment uses the GB1 fitness landscape
introduced by Wu et al.~\cite{wu2016}. Four positions in the B1 domain
of protein G are varied:

\begin{equation}
\mathrm{V39},\quad
\mathrm{D40},\quad
\mathrm{G41},\quad
\mathrm{V54}.
\end{equation}

The complete combinatorial space contains

\begin{equation}
20^4=160{,}000
\end{equation}

possible variants, of which 149,361 were experimentally measured.

GB1 is well suited to ORBIT because it contains substantial epistatic
structure while remaining small enough for exact order-resolved analysis
over complete local four-site landscapes.

\subsection{Primary FLIP 2-vs-Rest Split}

For the primary prediction experiment, we adopt the official FLIP GB1
2-vs-rest split \cite{dallago2021}. The model-training partition contains
wild type, single mutants, and double mutants,

\begin{equation}
\mathrm{HD}\leq 2,
\end{equation}

whereas the test partition contains triple and quadruple mutants,

\begin{equation}
\mathrm{HD}=3\ \text{or}\ 4.
\end{equation}

The released FLIP split artifact used in our experiments contains
424 training variants and 8,309 test variants. The FLIP test labels are
never used for model fitting, hyperparameter selection, or early-stopping
decisions.

This experiment therefore measures whether models trained only on
lower-mutation-complexity variants can generalize to experimentally
measured variants carrying three or four simultaneous mutations.

The FLIP split is the upstream model-training and prediction-evaluation
split. It is distinct from the AA-identity-strict probe split used later
for representation-accessibility analysis.

\subsection{WT-Anchored Local Binary Fitness Cubes}

For order-resolved analysis, we construct four-dimensional binary
fitness cubes anchored at the GB1 wild type

\begin{equation}
\mathrm{VDGV}.
\end{equation}

At each of the four positions, the binary states are

\begin{equation}
0 = \text{WT amino acid},
\qquad
1 = \text{one selected non-WT amino acid}.
\end{equation}

Selecting one non-WT alternative at each position defines a local
WT-anchored cube with

\begin{equation}
2^4=16
\end{equation}

variants: the WT sequence, four single mutants, six double mutants,
four triple mutants, and one quadruple mutant.

Because each position has 19 possible non-WT amino-acid alternatives,
there are

\begin{equation}
19^4=130{,}321
\end{equation}

candidate WT-anchored cubes.

A cube is retained only when all 16 vertices have experimentally
measured GB1 fitness values. Applying this completeness criterion to the
149,361 measured variants yields

\begin{equation}
109{,}235
\end{equation}

complete WT-anchored cubes.

These complete local landscapes provide the shared discrete objects used
for function-level Walsh decomposition and representation-level ORBIT
analysis.

\section{Models}

We use lightweight models so that the study isolates representation
organization and downstream nonlinear processing rather than effects of
large-scale pretraining. Five predictors are evaluated in the primary
GB1 experiment: ridge regression, a standard MLP, independent tokens,
nonlinear independent tokens, and Residual Interaction Tokenization
(RIT).

Ridge regression is included in prediction and functional-recovery
analyses but has no learned hidden representation for ORBIT probing.
The MLP is analyzed at hidden stages $H_1$ and $H_2$. The three token
models additionally expose an explicit token representation
$Z_{\mathrm{tok}}$ before the shared downstream backbone.

\subsection{Ridge Regression}

A ridge regression model on one-hot residue encodings serves as a linear
baseline:

\begin{equation}
\hat{y}
=
\mathbf{w}^{\top}x+b.
\end{equation}

Because the input contains position-specific residue indicators without
explicit cross-position interaction features, this model primarily
provides an additive reference for the nonlinear architectures.

\subsection{Standard MLP}

The standard MLP receives the concatenated one-hot sequence
representation and applies two nonlinear hidden layers:

\begin{align}
h_1 &= \sigma(W_1x+b_1),\\
h_2 &= \sigma(W_2h_1+b_2),\\
\hat{y} &= W_oh_2+b_o.
\end{align}

The primary MLP therefore has no explicit interaction-aware tokenization.
Any cross-position interaction structure must be constructed by the
downstream nonlinear transformations.

\subsection{Independent Token Representation}

For position $i$, an independently learned token representation is

\begin{equation}
u_i=E_i(x_i),
\end{equation}

where $E_i$ maps the residue identity at position $i$ to a learned token
vector.

The four token vectors are then concatenated:

\begin{equation}
z_{\mathrm{tok}}
=
u_1\Vert u_2\Vert u_3\Vert u_4.
\end{equation}

At this stage, each token depends only on its own residue identity.
Thus, $Z_{\mathrm{tok}}$ contains no explicit cross-position interaction
mechanism.

\subsection{Nonlinear Independent Tokens}

To distinguish token-wise nonlinear transformation from explicit
cross-position interaction modeling, we introduce an
architecture-matched mechanistic control:

\begin{equation}
u_i
=
\phi(E_i(x_i)).
\end{equation}

The nonlinear map $\phi$ is applied independently to each position.
Consequently, each token can undergo nonlinear transformation while
remaining independent of the residue identities at all other positions.

This control allows us to distinguish an effect caused by additional
token-level nonlinearity from one caused specifically by cross-position
interaction.

\subsection{Residual Interaction Tokenization}

RIT introduces explicit cross-position information before the shared
downstream predictor. Starting from position-specific token vectors
$u_i$, it computes

\begin{equation}
c_i
=
\frac{1}{d-1}
\sum_{j\neq i}
\phi_{\mathrm{int}}(u_i,u_j),
\end{equation}

and forms an interaction-aware residual token

\begin{equation}
z_i
=
u_i
+
\alpha\rho(u_i,c_i).
\label{eq:rit}
\end{equation}

The residual coefficient $\alpha$ is a learnable scalar initialized to
$0.05$, so the representation begins close to the independent-token
model and can learn the magnitude of the interaction-dependent
correction.

The sequence-level token representation is

\begin{equation}
z_{\mathrm{tok}}
=
z_1\Vert z_2\Vert z_3\Vert z_4.
\end{equation}

\paragraph{Interaction-order interpretation.}
The context term $c_i$ is constructed from pairwise functions
$\phi_{\mathrm{int}}(u_i,u_j)$. RIT therefore provides a direct
mechanism for introducing pairwise cross-position information at
$Z_{\mathrm{tok}}$.

This does not guarantee genuine third- or fourth-order accessibility.
Whether higher-order information emerges from subsequent composition of
these pairwise terms is an empirical question evaluated by ORBIT rather
than an architectural assumption.

\subsection{Matched Primary Downstream Backbone}

The independent-token, nonlinear-token, and RIT models share the same
downstream predictor. After construction of $Z_{\mathrm{tok}}$,

\begin{align}
h_1
&=
\sigma(W_1 z_{\mathrm{tok}}+b_1),\\
h_2
&=
\sigma(W_2 h_1+b_2),\\
\hat{y}
&=
W_o h_2+b_o.
\end{align}

The token dimension, hidden dimensions, optimizer, training budget, and
FLIP data partitions are matched across these three primary
architectures. Their principal architectural difference is therefore
how $Z_{\mathrm{tok}}$ is constructed before entering the common
two-hidden-layer backbone.

This matched-backbone comparison is the primary architecture experiment.

\subsection{Pre-Specified Depth/Capacity Sensitivity Analysis}

We additionally evaluate whether the primary conclusions are sensitive
to a deeper downstream predictor.

Before inspecting the depth results, we froze a confirmatory analysis
covering neural backbones with

\begin{equation}
d_{\mathrm{hidden}}\in\{2,3,4\}.
\end{equation}

Depth 2 corresponds to the primary architecture described above.
Depths 3 and 4 append additional hidden transformations after $H_2$,
producing

\begin{equation}
H_1
\rightarrow
H_2
\rightarrow
H_3
\end{equation}

and

\begin{equation}
H_1
\rightarrow
H_2
\rightarrow
H_3
\rightarrow
H_4,
\end{equation}

respectively.

The deeper variants are evaluated for the MLP, independent tokens,
nonlinear independent tokens, and RIT using the same 20 paired training
seeds as the primary analysis. The pre-specified confirmatory contrasts
are

\begin{equation}
\text{depth 3} - \text{depth 2}
\end{equation}

and

\begin{equation}
\text{depth 4} - \text{depth 2}.
\end{equation}

Depth 4 versus depth 3 is not treated as a confirmatory comparison.

For representation accessibility, the final layer is defined as $H_2$
for depth 2, $H_3$ for depth 3, and $H_4$ for depth 4.

Because the deeper networks contain additional layers and therefore
additional parameters, this experiment is interpreted as a
\emph{depth/capacity sensitivity analysis}. It does not isolate a causal
effect of depth under parameter-matched capacity.

\section{Evaluation and Statistical Inference}

We evaluate the models using complementary prediction, functional,
and representation-level measurements.

\subsection{Prediction Metrics}

On the official FLIP test partition, we report

\begin{equation}
\mathrm{RMSE},
\qquad
R^2,
\qquad
\rho_{\mathrm{Spearman}}.
\end{equation}

These metrics summarize overall generalization from the FLIP
model-training partition ($\mathrm{HD}\leq 2$) to experimentally
measured triple and quadruple mutants ($\mathrm{HD}=3$ or $4$).

FLIP test $R^2$ is the pre-specified confirmatory prediction endpoint.
RMSE and Spearman correlation are reported descriptively to provide
complementary views of predictive performance.

\subsection{Functional Epistasis Recovery}

For each interaction order $k\in\{1,2,3,4\}$, we report

\begin{equation}
R_k^2,
\end{equation}

which measures agreement between experimentally measured and predicted
order-specific Walsh coefficients.

The higher-order confirmatory endpoints are

\begin{equation}
R_3^2
\qquad \text{and} \qquad
R_4^2.
\end{equation}

First- and second-order recovery are also reported but are not included
in the higher-order confirmatory hypothesis family.

\subsection{Representation Walsh Energy}

For each representation stage $\ell$ and interaction order $k$, we
report normalized representation Walsh energy

\begin{equation}
E_{\ell,k}.
\end{equation}

This quantity measures the fraction of nonconstant representation Walsh
energy associated with order $k$.

Representation energy is treated as a descriptive quantity throughout
the study. We do not perform heatmap-wide inferential testing on
$E_{\ell,k}$.

\subsection{Representation Accessibility}

For each representation stage $\ell$ and interaction order $k$, we
report

\begin{equation}
A_{\ell,k},
\end{equation}

defined as held-out ridge-probe $R^2$ for predicting experimentally
measured order-$k$ Walsh coefficients from representation-level Walsh
components.

For the primary depth-2 architecture comparison, the confirmatory
higher-order accessibility endpoints are

\begin{equation}
A_{H_2,3}
\qquad \text{and} \qquad
A_{H_2,4}.
\end{equation}

For the depth/capacity sensitivity analysis, the corresponding
confirmatory endpoints are final-layer

\begin{equation}
A_3
\qquad \text{and} \qquad
A_4,
\end{equation}

where the final representation is $H_2$, $H_3$, or $H_4$ for depths
2, 3, and 4, respectively.

We additionally test token-stage pairwise accessibility

\begin{equation}
A_{\mathrm{tok},2}
\end{equation}

for RIT relative to the two independent-token controls.

Intermediate-layer accessibility trajectories are reported
descriptively unless they correspond to one of these pre-specified
confirmatory endpoints.

\subsection{Complementary ORBIT Views}

The ORBIT measurements are intentionally not collapsed into a single
score. Instead, we distinguish

\begin{equation}
\boxed{
\text{presence}
\;\longrightarrow\;
\text{accessibility}
\;\longrightarrow\;
\text{functional recovery}
}
\end{equation}

corresponding respectively to

\begin{equation}
E_{\ell,k},
\qquad
A_{\ell,k},
\qquad
R_k^2.
\end{equation}

These measurements address different questions: whether order-specific
variation exists in a representation, whether experimentally measured
interaction structure can be linearly recovered from that
representation, and whether it is ultimately reproduced by the model's
predicted fitness landscape.

\subsection{Confirmatory Statistical Inference}

All stochastic neural-model inference uses the same 20 paired training
seeds,

\begin{equation}
42,\ldots,61.
\end{equation}

Ridge regression is deterministic and is therefore reported
descriptively rather than treated as a stochastic replicate in paired
seed-based inference.

For each confirmatory neural-model comparison, we use a two-sided exact
paired sign-flip permutation test over all

\begin{equation}
2^{20}=1{,}048{,}576
\end{equation}

sign assignments.

For each comparison we report:

\begin{itemize}[leftmargin=*,itemsep=1pt]
    \item the mean paired difference;
    \item a 95\% paired percentile-bootstrap confidence interval;
    \item paired standardized effect size $d_z$;
    \item the raw exact permutation $p$-value; and
    \item the Holm-adjusted $p$-value.
\end{itemize}

Bootstrap confidence intervals use 20,000 paired resamples.

\subsection{Pre-Specified Multiplicity Families}

The primary architecture analysis originally contained 17 confirmatory
tests. Before inspection of the depth/capacity results, the hypothesis
plan was frozen and 40 depth comparisons were added, yielding

\begin{equation}
57
\end{equation}

confirmatory tests in total.

Holm adjustment is applied separately within four pre-specified
families:

\begin{enumerate}[leftmargin=*,itemsep=1pt]

    \item \textbf{Prediction:}
    FLIP test $R^2$,
    11 tests total;

    \item \textbf{Higher-order functional recovery:}
    $R_3^2$ and $R_4^2$,
    22 tests total;

    \item \textbf{Final-layer higher-order accessibility:}
    $A_3$ and $A_4$,
    22 tests total;

    \item \textbf{Token-stage pairwise accessibility:}
    $A_{\mathrm{tok},2}$,
    2 tests total.

\end{enumerate}

Statistical significance is defined as

\begin{equation}
p_{\mathrm{Holm}}\leq 0.05.
\end{equation}

The pre-specified depth contrasts are depth 3 minus depth 2 and depth 4
minus depth 2 for each neural architecture. Depth 4 minus depth 3 is not
a confirmatory comparison.

All pre-specified depth results are reported regardless of statistical
significance or direction. Representation energy and intermediate-layer
trajectories remain descriptive and are not used to define additional
post hoc confirmatory hypotheses.
Complete reproducibility details, including the frozen execution order,
paired-seed design, probe-fitting protocol, and analysis provenance, are
provided in Appendix~\ref{app:reproducibility}.
\section{Experimental Questions}

\paragraph{Q1: Does ORBIT correctly recover known interaction order?}

We first validate the Walsh-based ORBIT diagnostics on synthetic
landscapes with known first- through fourth-order structure, exact
fourth-order parity, and a controlled transition from lower- to
higher-order Walsh energy.

\paragraph{Q2: Can models trained on lower-mutation-complexity variants
generalize to triple and quadruple mutants?}

Under the official FLIP 2-vs-rest protocol, models are trained only on
wild type, single mutants, and double mutants and are evaluated on
experimentally measured triple and quadruple mutants.

\paragraph{Q3: Where does order-specific information become accessible
inside learned representations?}

For each learned representation stage $\ell$, we measure normalized
representation Walsh energy

\begin{equation}
E_{\ell,k}
\end{equation}

and linear accessibility

\begin{equation}
A_{\ell,k}
\end{equation}

for interaction orders $k=1,\ldots,4$.

\paragraph{Q4: Does interaction-aware tokenization alter the point at
which interaction information first becomes accessible?}

In the primary matched-backbone comparison, we compare Independent
tokens, Nonlinear independent tokens, and RIT at

\begin{equation}
Z_{\mathrm{tok}},
\qquad
H_1,
\qquad
H_2.
\end{equation}

The central token-stage test asks whether RIT increases pairwise
accessibility $A_{\mathrm{tok},2}$ relative to the two independent-token
controls.

\paragraph{Q5: Can predictive performance, representation accessibility,
and functional recovery diverge?}

We compare conventional fitness-prediction performance with
order-specific accessibility and functional recovery to determine
whether models with similar final prediction performance can organize
interaction information differently internally.

\paragraph{Q6: Are the conclusions sensitive to additional downstream
depth and capacity?}

Under the pre-result frozen depth/capacity analysis, we compare neural
backbones with two, three, and four hidden layers.

The pre-specified confirmatory contrasts are

\begin{equation}
\text{depth 3} - \text{depth 2}
\end{equation}

and

\begin{equation}
\text{depth 4} - \text{depth 2},
\end{equation}

evaluated for FLIP test $R^2$, strict-probe $R_3^2$ and $R_4^2$, and
final-layer $A_3$ and $A_4$.

\section{Results}

\subsection{ORBIT-Synth Validation}

ORBIT recovered the interaction order planted in each controlled
synthetic landscape. In the pure-order benchmarks, normalized
nonconstant Walsh energy was concentrated entirely at the corresponding
first-, second-, third-, or fourth-order component. The exact parity
control likewise concentrated all nonconstant energy at order 4.

The controlled cooperativity sweep recovered the prescribed transition
from first- to fourth-order structure. Because the parameter $C$
specifies the fraction of nonconstant energy assigned to the
higher-order component, the recovered spectra followed

\begin{equation}
E_1 = 1-C,
\qquad
E_4 = C,
\end{equation}

with zero energy at orders 2 and 3.

Thus, increasing $C$ from 0 to 1 produced the intended progression from
purely first-order to purely fourth-order structure. These results
confirm that the Walsh-based ORBIT diagnostics recover known interaction
order under controlled ground truth before application to experimental
GB1 data.
\subsection{GB1 Fitness Generalization}

\begin{table}[htbp]
\centering
\caption{Prediction performance on the official FLIP \texttt{two\_vs\_rest} test set (Hamming distance 3--4). Neural-model values are mean $\pm$ standard deviation across 20 seeds (42--61); Ridge is deterministic under the frozen configuration.}
\label{tab:gb1_prediction}
\begin{tabular}{lccc}
\toprule
Model & RMSE $\downarrow$ & $R^2$ $\uparrow$ & Spearman $\rho$ $\uparrow$ \\
\midrule
Ridge & 1.1619 ± 0.0000 & 0.1084 ± 0.0000 & 0.5617 ± 0.0000 \\
MLP & 1.1377 ± 0.0986 & 0.1392 ± 0.1531 & 0.6667 ± 0.0208 \\
Independent tokens & 1.0601 ± 0.0604 & 0.2556 ± 0.0862 & 0.6054 ± 0.0365 \\
Nonlinear tokens & 1.0898 ± 0.0876 & 0.2109 ± 0.1298 & 0.7066 ± 0.0210 \\
RIT & 1.0746 ± 0.0737 & 0.2340 ± 0.1085 & 0.6037 ± 0.0344 \\
\bottomrule
\end{tabular}
\end{table}

All nonlinear neural models generalized beyond the mutation complexity
represented in the FLIP training partition, although none emerged as a
statistically dominant architecture in the primary two-hidden-layer
comparison.

Descriptively, the Independent-token model achieved the highest mean
FLIP test $R^2$ and lowest mean RMSE, whereas the Nonlinear-token model
achieved the highest mean Spearman correlation. RIT remained competitive
with both controls.

However, the pre-specified seed-paired comparisons found no significant
differences in FLIP test $R^2$ among the token architectures after Holm
correction. The RIT--Independent, RIT--Nonlinear, and
Nonlinear--Independent contrasts were all nonsignificant.

Thus, explicit pairwise interaction-aware tokenization did not provide
a detectable advantage in final GB1 fitness generalization under the
primary matched-backbone setting.
\subsection{Order-Specific Epistasis Recovery}

\begin{figure}[H]
    \centering
    \includegraphics[width=0.70\linewidth]{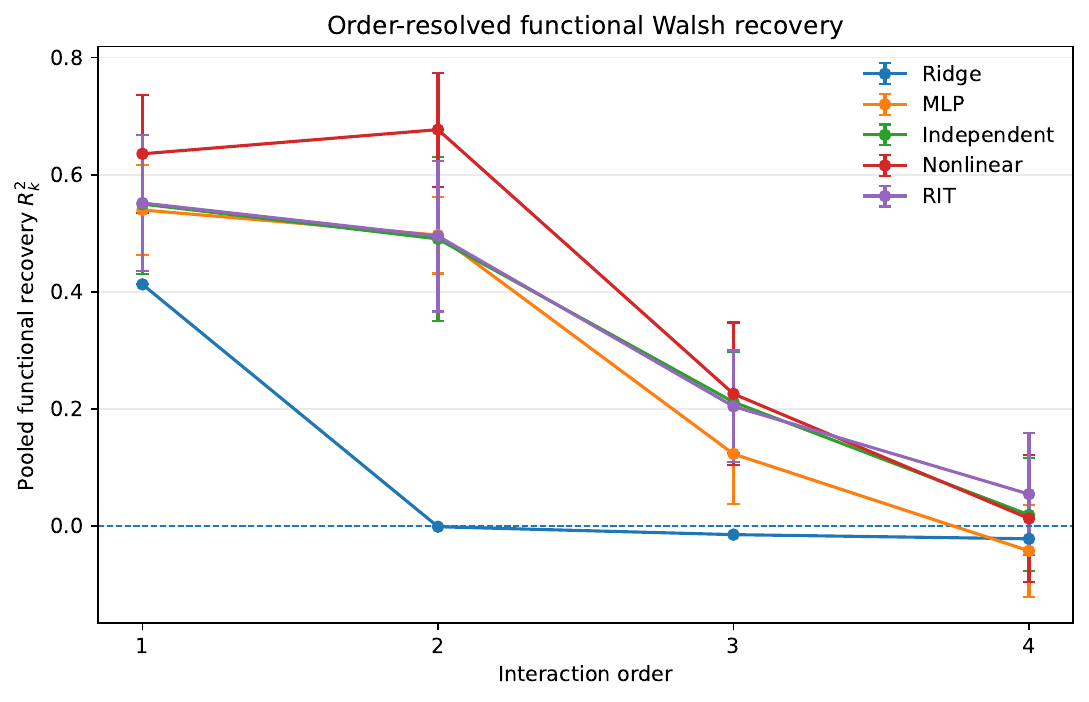}
    \caption{
    \textbf{Order-resolved functional Walsh recovery on the
    AA-identity-strict probe test.}
    Pooled functional recovery $R_k^2$ is shown for interaction
    orders $k=1,\ldots,4$.
    Neural-model points indicate means across 20 stochastic
    realizations (seeds 42--61), with error bars showing
    $\pm 1$ standard deviation. Ridge is deterministic.
    Recovery decreases as interaction order increases, with
    substantially greater uncertainty at third and fourth order.
    }
    \label{fig:functional_walsh_recovery}
\end{figure}

As shown in Fig.~\ref{fig:functional_walsh_recovery},
order-resolved analysis revealed a progressive decline in functional
recovery as interaction order increased.

The nonlinear neural models
recovered first- and second-order experimental Walsh structure
substantially better than the linear ridge baseline, whereas recovery
became weaker and more variable at third and especially fourth order.

Descriptively, the Nonlinear-token model achieved the highest mean
functional recovery through third order, while RIT achieved the highest
mean fourth-order recovery. These rankings, however, were not supported
by significant architecture-level differences at the higher orders.

For the pre-specified third- and fourth-order endpoints, none of the
RIT--Independent, RIT--Nonlinear, or Nonlinear--Independent contrasts
was significant after Holm correction.

Thus, the primary matched-backbone experiment provides no statistical
evidence that interaction-aware tokenization improves final
third- or fourth-order functional recovery. The descriptive differences
at these orders should therefore not be interpreted as reliable
architectural advantages.

Complete order-resolved functional-recovery values for all primary
models are reported in Appendix~\ref{app:complete_primary_results}.
\subsection{Representation Presence and Accessibility}

\begin{figure}[H]
    \centering

    \begin{minipage}[t]{0.48\linewidth}
        \centering
        \includegraphics[width=\linewidth]{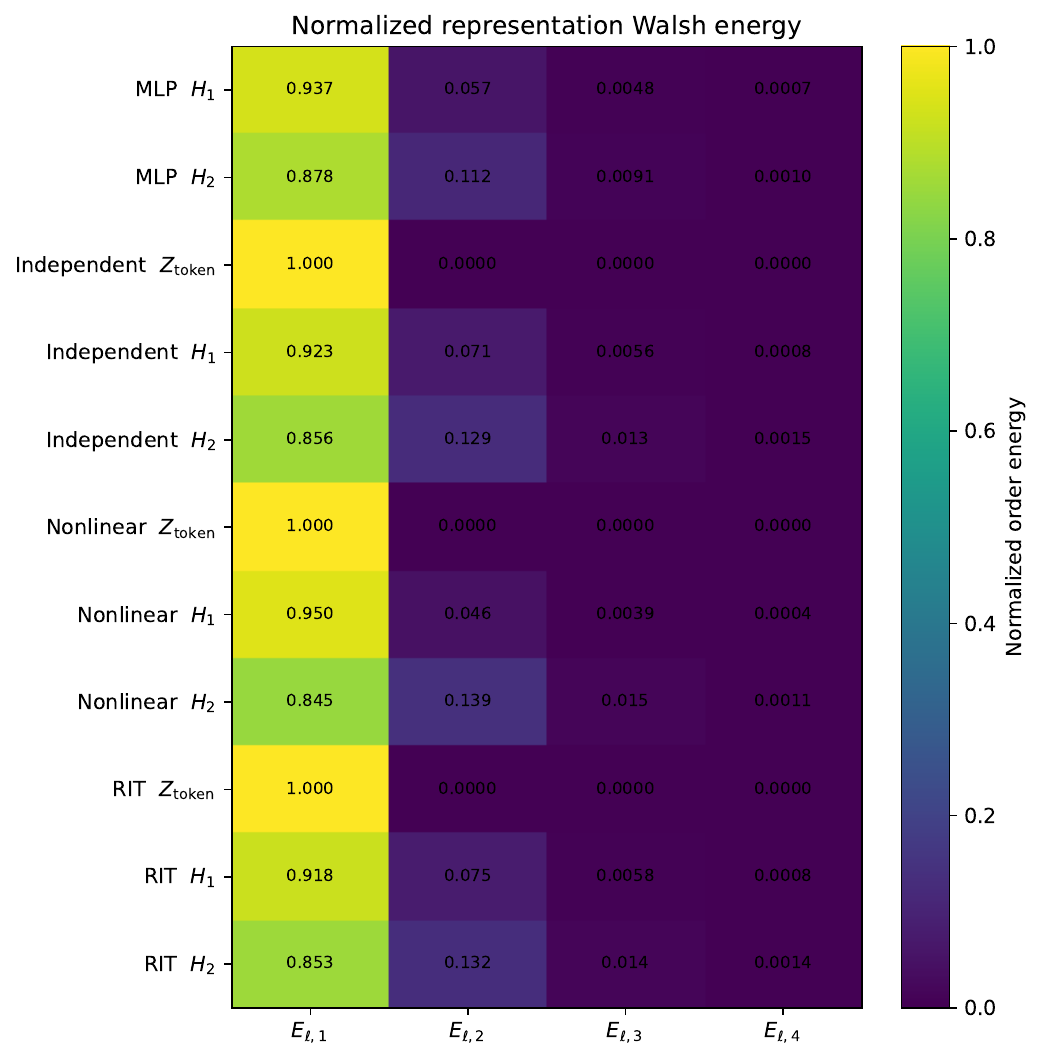}

        \small (A) Representation Walsh energy
    \end{minipage}
    \hfill
    \begin{minipage}[t]{0.48\linewidth}
        \centering
        \includegraphics[width=\linewidth]{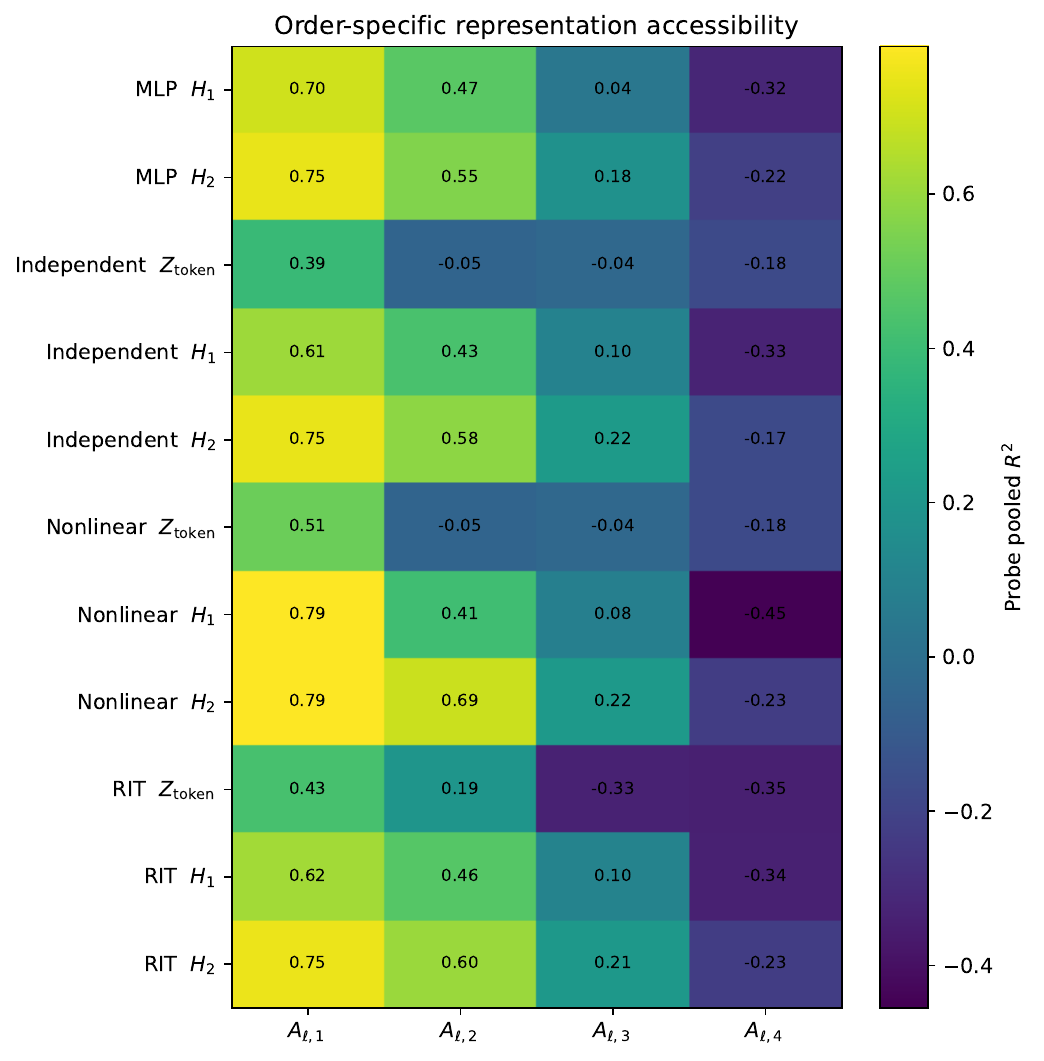}

        \small (B) Representation accessibility
    \end{minipage}

    \caption{
    \textbf{Order-specific structure across learned GB1 representations.}
    (A) Normalized representation Walsh energy $E_{\ell,k}$ quantifies
    the fraction of nonconstant representation variation associated
    with interaction order $k$.
    (B) Accessibility $A_{\ell,k}$ measures held-out linear-probe
    recovery of experimentally measured order-$k$ Walsh coefficients
    from each representation stage.
    Energy and accessibility capture distinct properties: the presence
    of order-specific variation does not imply that experimentally
    measured epistatic structure is linearly accessible.
    }
    \label{fig:representation_orbit}
\end{figure}

As shown in Fig.~\ref{fig:representation_orbit},
representation-level ORBIT analysis revealed clear differences between
the presence of order-specific variation and its accessibility for
recovering experimentally measured epistatic structure.

Normalized representation Walsh energy was dominated by first-order
components across architectures. At the token stage, the Independent
and Nonlinear-independent models contained only first-order structure by
construction, whereas RIT introduced additional second-order
representation energy through its explicit pairwise interaction
mechanism. Subsequent hidden layers redistributed some representation
energy toward higher orders, although third- and fourth-order energy
remained comparatively small.

Accessibility produced a stronger architecture-specific distinction.
At the token stage, RIT substantially increased pairwise accessibility
relative to both independent-token controls. The paired difference was

\begin{equation}
\Delta A_{\mathrm{tok},2}=0.2468,
\end{equation}

with

\begin{equation}
d_z=1.67,
\qquad
p_{\mathrm{Holm}}=1.14\times10^{-5}.
\end{equation}

The effect was identical relative to the Independent-token and
Nonlinear-independent-token controls and remained significant after
correction within the pre-specified token-stage family.

This early pairwise advantage did not extend to the pre-specified
higher-order final-layer endpoints. None of the architecture contrasts
for final-layer $A_3$ or $A_4$ was significant after Holm correction.

Thus, RIT changes where pairwise information first becomes linearly
accessible, but the primary two-hidden-layer experiment provides no
evidence that this early representational change produces a reliable
third- or fourth-order accessibility advantage at the final hidden
layer.

Complete stage- and order-resolved accessibility and representation
Walsh-energy values are reported in
Appendix~\ref{app:complete_primary_results}.

\subsection{Depth/Capacity Sensitivity Analysis}

We next evaluated whether the primary conclusions were sensitive to
additional downstream depth and capacity. The analysis was
pre-specified before inspection of the depth results and compared
two-hidden-layer backbones with corresponding three- and
four-hidden-layer variants.

Complete numerical depth/capacity results are reported in
Appendix~\ref{app:depth_results}.

\begin{figure}[H]
    \centering

    \begin{minipage}[t]{0.48\linewidth}
        \centering
        \includegraphics[width=\linewidth]
        {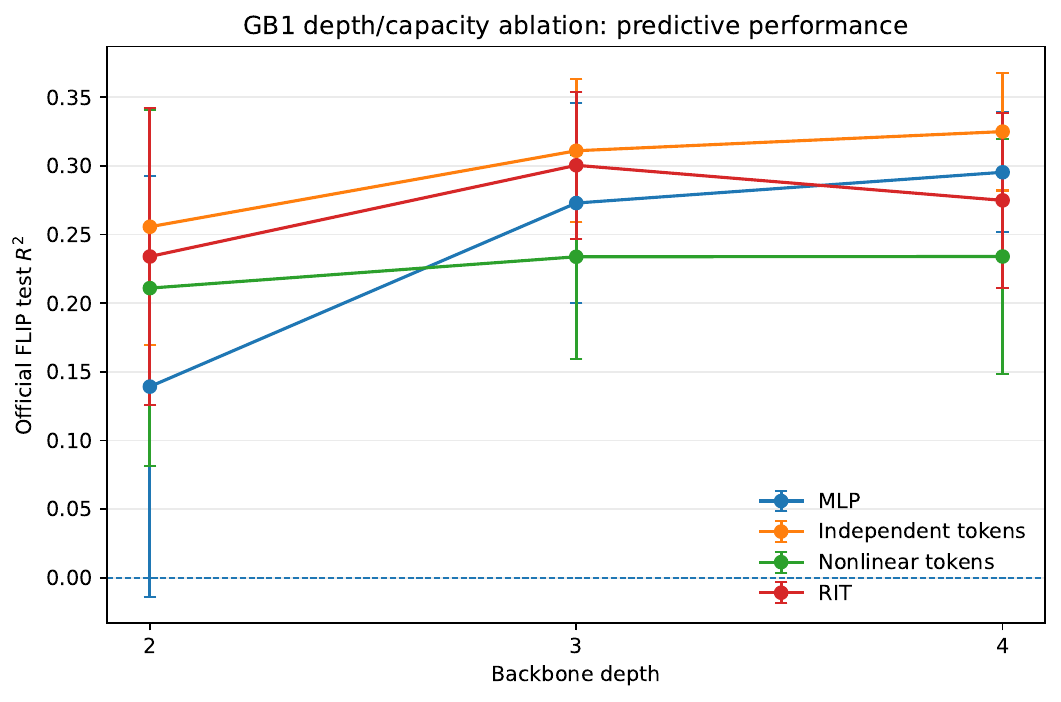}

        \small (A) FLIP prediction
    \end{minipage}
    \hfill
    \begin{minipage}[t]{0.48\linewidth}
        \centering
        \includegraphics[width=\linewidth]
        {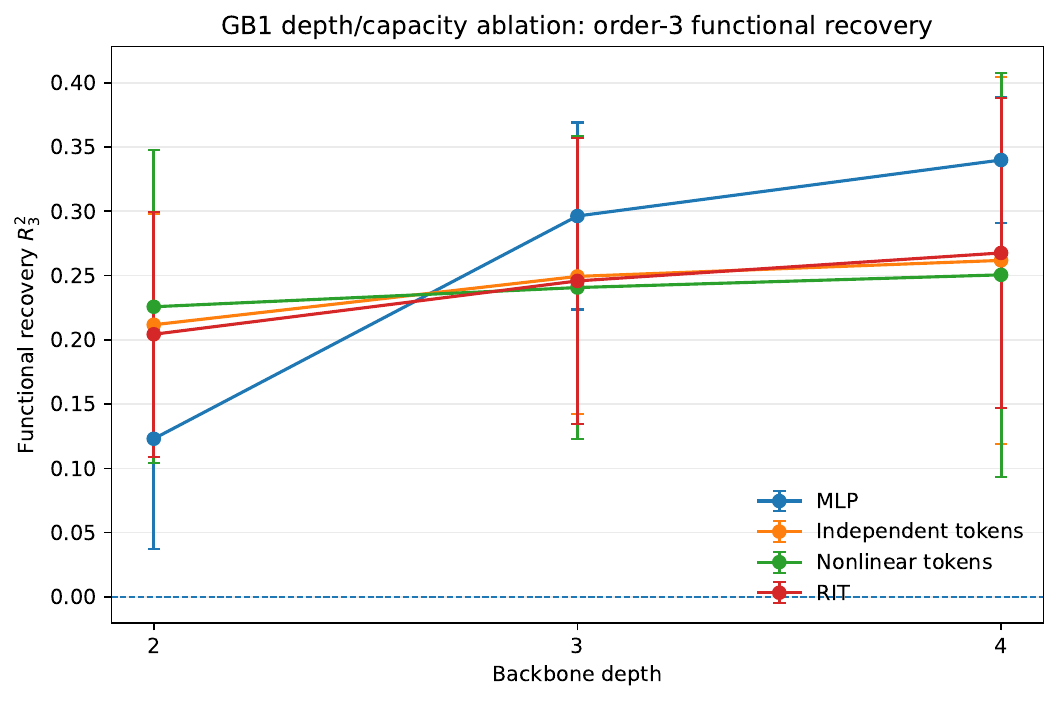}

        \small (B) Third-order functional recovery
    \end{minipage}

    \caption{
    \textbf{Depth/capacity sensitivity of GB1 prediction and
    third-order functional recovery.}
    Neural backbones with two, three, and four hidden layers are
    evaluated using the same 20 paired training seeds.
    (A) Official FLIP test $R^2$.
    (B) Strict-probe third-order functional recovery $R_3^2$.
    The standard MLP shows the clearest depth-associated gains in both
    prediction and third-order recovery. Because deeper networks also
    contain additional parameters, these results are interpreted as
    depth/capacity sensitivity rather than a parameter-matched causal
    effect of depth.
    }
    \label{fig:depth_sensitivity}
\end{figure}

\paragraph{Prediction.}

As shown in Fig.~\ref{fig:depth_sensitivity}, the clearest prediction
effect occurred for the standard MLP.
Relative to depth 2, FLIP test $R^2$ increased at both depth 3

\begin{equation}
\Delta R^2 = 0.134,
\qquad
d_z = 0.88,
\qquad
p_{\mathrm{Holm}} = 0.0057,
\end{equation}

and depth 4

\begin{equation}
\Delta R^2 = 0.156,
\qquad
d_z = 1.08,
\qquad
p_{\mathrm{Holm}} = 6.7\times10^{-4}.
\end{equation}

Independent tokens also showed a significant prediction improvement at
depth 4

\begin{equation}
\Delta R^2 = 0.069,
\qquad
d_z = 0.82,
\qquad
p_{\mathrm{Holm}} = 0.0148,
\end{equation}

whereas its depth-3 contrast did not survive multiplicity correction.
Neither Nonlinear-independent tokens nor RIT showed a Holm-significant
confirmatory depth effect on FLIP test $R^2$.

\paragraph{Functional recovery.}

Additional MLP depth also substantially improved third-order functional
recovery. Relative to depth 2,

\begin{equation}
\Delta R_3^2 = 0.173
\end{equation}

at depth 3 and

\begin{equation}
\Delta R_3^2 = 0.217
\end{equation}

at depth 4, with both contrasts significant after Holm correction

\begin{equation}
p_{\mathrm{Holm}} = 4.20\times10^{-5}.
\end{equation}

No corresponding Holm-significant third-order depth effect was detected
for Independent tokens, Nonlinear-independent tokens, or RIT.

None of the four architectures showed a significant confirmatory depth
effect on fourth-order functional recovery $R_4^2$ after Holm
correction.

\paragraph{Final-layer accessibility.}

The MLP likewise showed increased final-layer third-order accessibility.
Relative to depth 2,

\begin{equation}
\Delta A_3 = 0.106
\end{equation}

at depth 3 and

\begin{equation}
\Delta A_3 = 0.130
\end{equation}

at depth 4, with Holm-adjusted $p$-values of $0.0015$ and
$7.13\times10^{-4}$, respectively.

Fourth-order accessibility also improved for the deeper MLP:

\begin{equation}
\Delta A_4 = 0.165
\end{equation}

at depth 3 and

\begin{equation}
\Delta A_4 = 0.177
\end{equation}

at depth 4.

Both contrasts were significant after Holm correction. Importantly,
however, the absolute held-out fourth-order accessibility remained below
zero at every MLP depth. The result therefore indicates a significant
improvement relative to the shallow MLP, not successful absolute
fourth-order decoding.

No token-based architecture showed a Holm-significant confirmatory depth
effect on final-layer $A_3$ or $A_4$.

\paragraph{Interpretation.}

These results reveal a second route by which interaction structure can
emerge. RIT changes pairwise accessibility directly at the token stage,
whereas additional downstream MLP capacity can construct substantially
more third-order functional and representation-level structure later in
the network.

The depth experiment does not establish a parameter-matched causal
effect of depth, because the deeper networks also contain additional
parameters. We therefore interpret these findings as
\emph{depth/capacity sensitivity}: downstream nonlinear capacity can
alter higher-order recovery even when the initial representation is not
explicitly interaction-aware.

\subsection{Prediction and Representation Can Diverge}

\begin{table}[htbp]
\centering
\caption{Primary two-hidden-layer architecture analysis before the
pre-specified depth/capacity extension. Paired statistical comparisons
are based on 20 stochastic training seeds. Differences are defined as
Model A $-$ Model B. Confidence intervals are paired-bootstrap 95\%
intervals, and $d_z$ denotes the standardized paired effect size.
$p_{\mathrm{Holm}}$ values in this table correspond to the original
primary-analysis multiplicity families. The subsequently frozen
depth/capacity extension and the final 57-test multiplicity analysis
are reported in Appendix~\ref{app:complete_statistics}.}
\label{tab:gb1_statistical_tests}
\small
\setlength{\tabcolsep}{4.5pt}
\begin{tabular}{llrrrr}
\toprule
Endpoint & Comparison & $\Delta$ & 95\% CI & $d_z$ & $p_{\mathrm{Holm}}$ \\
\midrule
FLIP test $R^2$ & RIT $-$ Independent & -0.022 & [-0.066, 0.019] & -0.222 & 0.687 \\
FLIP test $R^2$ & RIT $-$ Nonlinear & 0.023 & [-0.049, 0.097] & 0.136 & 0.687 \\
\addlinespace[2pt]
Functional $R_3^2$ & RIT $-$ Independent & -0.007 & [-0.046, 0.031] & -0.082 & 1.000 \\
Functional $R_3^2$ & RIT $-$ Nonlinear & -0.021 & [-0.079, 0.038] & -0.156 & 1.000 \\
\addlinespace[2pt]
Functional $R_4^2$ & RIT $-$ Independent & 0.036 & [-0.007, 0.077] & 0.360 & 0.741 \\
Functional $R_4^2$ & RIT $-$ Nonlinear & 0.042 & [-0.019, 0.101] & 0.301 & 0.983 \\
\addlinespace[2pt]
$H_2$ accessibility $A_3$ & RIT $-$ Independent & -0.015 & [-0.055, 0.022] & -0.168 & 1.000 \\
$H_2$ accessibility $A_3$ & RIT $-$ Nonlinear & -0.009 & [-0.063, 0.050] & -0.064 & 1.000 \\
\addlinespace[2pt]
$H_2$ accessibility $A_4$ & RIT $-$ Independent & -0.058 & [-0.158, 0.046] & -0.240 & 1.000 \\
$H_2$ accessibility $A_4$ & RIT $-$ Nonlinear & -0.003 & [-0.125, 0.119] & -0.010 & 1.000 \\
\addlinespace[2pt]
\textbf{$Z_{\mathrm{tok}}$ accessibility $A_2$} & \textbf{RIT $-$ Independent} & \textbf{0.247} & \textbf{[0.183, 0.309]} & \textbf{1.673} & \textbf{$1.14\times10^{-5}$} \\
\textbf{$Z_{\mathrm{tok}}$ accessibility $A_2$} & \textbf{RIT $-$ Nonlinear} & \textbf{0.247} & \textbf{[0.182, 0.309]} & \textbf{1.673} & \textbf{$1.14\times10^{-5}$} \\
\bottomrule
\end{tabular}
\end{table}

The primary-analysis conclusions are unchanged under the subsequently
expanded 57-test confirmatory analysis; complete final-family-adjusted
results are reported in Appendix~\ref{app:complete_statistics}.

Taken together, the primary architecture and depth/capacity analyses
show that conventional fitness prediction does not fully determine how
interaction information is organized internally.

In the primary matched-backbone comparison, the token architectures
showed no significant differences in FLIP test $R^2$, third- or
fourth-order functional recovery, or final-layer $A_3$ and $A_4$.
Nevertheless, RIT produced a large and highly significant increase in
pairwise accessibility directly at the token stage.

Conversely, the depth/capacity analysis showed that later nonlinear
processing can substantially alter higher-order structure without
changing the initial tokenization mechanism. In particular, deeper MLPs
improved FLIP prediction, third-order functional recovery, and
final-layer third- and fourth-order accessibility, while no comparable
confirmatory depth effect was detected for RIT on these higher-order
endpoints.

These findings distinguish two representation pathways:

\begin{equation}
\boxed{
\text{early interaction encoding}
\quad \neq \quad
\text{later interaction construction}
}
\end{equation}

RIT provides evidence for the first pathway by exposing pairwise
information earlier in the representation trajectory. The deeper MLP
provides evidence for the second by showing that additional downstream
nonlinear capacity can construct more accessible higher-order structure
from an initially non-interaction-aware representation.

ORBIT therefore reveals differences that would be obscured by comparing
final prediction metrics alone.
\section{Discussion}

ORBIT was designed to separate questions that are easily conflated in
protein fitness modeling: whether order-specific structure is present
in a representation, whether experimentally measured epistatic
structure is linearly accessible from that representation, and whether
it is ultimately recovered in the predicted fitness landscape.
The ORBIT-Synth experiments establish that the Walsh-based diagnostics
recover known interaction order under controlled ground truth before
they are applied to experimental GB1 data.

The GB1 results show why this distinction matters. In the primary
matched-backbone comparison, no single token architecture was
consistently superior according to conventional fitness prediction.
Independent tokens achieved the highest mean FLIP test $R^2$ and the
lowest mean RMSE, whereas Nonlinear tokens achieved the highest mean
Spearman correlation. However, the pre-specified paired comparisons
provided no statistical evidence of differences in FLIP test $R^2$
among the token architectures. Thus, final prediction performance alone
does not reveal how these models organize epistatic information
internally.

The same pattern appears in functional recovery. All nonlinear models
recovered substantially more first- and second-order structure than the
linear ridge baseline, while recovery weakened as interaction order
increased. Descriptively, Nonlinear tokens achieved the strongest mean
recovery through third order and RIT the strongest mean fourth-order
recovery. Yet none of the pre-specified architecture contrasts for
$R_3^2$ or $R_4^2$ remained significant after multiplicity correction.
These differences should therefore be interpreted descriptively rather
than as evidence that one tokenization strategy reliably improves
higher-order functional recovery.

The clearest architecture-specific result instead occurs within the
representation trajectory. RIT substantially increased pairwise
accessibility directly at $Z_{\mathrm{tok}}$ relative to both
independent-token controls,

\begin{equation}
\Delta A_{\mathrm{tok},2}=0.2468,
\qquad
d_z=1.67,
\qquad
p_{\mathrm{Holm}}=1.14\times10^{-5}.
\end{equation}

This effect is consistent with the architecture: RIT explicitly mixes
pairwise cross-position information before the shared downstream
predictor. Importantly, the effect concerns \emph{accessibility}, not
merely representation energy. Although RIT introduces only a very small
fraction of second-order Walsh energy at the token stage, that component
is substantially better aligned with experimentally measured pairwise
epistasis than the corresponding components of the independent-token
controls.

The early RIT advantage did not propagate into a detectable
third- or fourth-order advantage at the final hidden layer, nor into
higher-order functional recovery or FLIP prediction. This distinction
shows that making interaction information accessible earlier is not
equivalent to improving the final learned function. An
interaction-aware representation can change \emph{where} information is
available without necessarily changing what the downstream predictor
ultimately reconstructs.

The depth/capacity experiment reveals the complementary phenomenon.
Increasing the standard MLP from two hidden layers to three or four
produced significant improvements in FLIP prediction, third-order
functional recovery, and final-layer third-order accessibility.
Fourth-order accessibility also improved significantly relative to the
two-layer MLP, although absolute $A_4$ remained below zero. Thus, the
deeper MLP became substantially better at exposing higher-order
structure without achieving reliable absolute fourth-order decoding.

These results suggest two distinct computational routes by which
interaction information can become accessible:

\begin{equation}
\boxed{
\text{early interaction encoding}
\quad\text{and}\quad
\text{later interaction construction}
}
\end{equation}

RIT provides evidence for the first route by introducing pairwise
cross-position structure directly into the token representation.
The deeper MLP provides evidence for the second: even when the initial
representation contains no explicit interaction mechanism, downstream
nonlinear transformations can progressively construct structure that is
more predictive of experimentally measured higher-order epistasis.

The distinction is particularly informative because the deeper RIT and
Nonlinear-token models did not show corresponding Holm-significant
improvements on the confirmatory higher-order endpoints. Additional
capacity therefore does not produce the same effect uniformly across
architectures. The interaction between representation design and
downstream processing appears to matter, rather than depth serving as a
universal solution.

At the same time, the depth experiment should not be interpreted as a
parameter-controlled causal demonstration that depth itself produces
higher-order structure. The three- and four-layer models contain more
parameters than their two-layer counterparts. The result therefore
establishes \emph{depth/capacity sensitivity}: increasing downstream
nonlinear capacity can alter higher-order recovery. Distinguishing the
specific contribution of depth from parameter count will require a
future parameter-matched comparison.

The FLIP 2-vs-rest setting further strengthens the interpretation of the
third-order results. Models are trained only on wild type, single
mutants, and double mutants, while prediction is evaluated on triple and
quadruple mutants. Positive third-order functional recovery therefore
shows that the nonlinear models can reconstruct useful order-specific
structure beyond the mutation complexity explicitly represented in the
training partition. Fourth-order recovery remains substantially more
difficult and shows no significant depth effect in the confirmatory
analysis.

Taken together, the experiments support a representation-centric view
of higher-order molecular learning. Models with similar predictive
performance can expose interaction information at different stages, and
models with similar initial representations can diverge after additional
nonlinear processing. ORBIT makes these differences measurable by
separating representation presence, linear accessibility, functional
recovery, and aggregate predictive performance rather than treating
them as interchangeable indicators of successful learning.
\section{Limitations and Future Work}

Several limitations define the scope of the present study.

First, GB1 varies only four amino-acid positions and therefore provides
a compact experimental system for studying epistasis rather than the
full complexity of protein sequence--function relationships. ORBIT
should next be evaluated on larger combinatorial landscapes and
additional molecular domains where interaction order can still be
defined or approximated reliably.

Second, Walsh coefficients quantify statistical epistasis. They should
not be interpreted automatically as direct physical contacts,
mechanistic biochemical interactions, or causal residue relationships.

Third, representation accessibility depends on the probe class. We use
a fixed ridge probe across architectures so that comparisons measure
linear accessibility under a common decoding rule. A nonlinear probe
could recover additional information, but would answer a different
question about decodability rather than linear accessibility.

Fourth, WT-anchored local cubes necessarily share the WT sequence and
can overlap in other sequence content. The AA-identity-strict probe
protocol addresses this by holding out non-WT amino-acid identities at
each position, but it is not an upstream sequence-disjoint
model-training split.

Fifth, the depth experiment changes both network depth and parameter
count. The significant improvements observed for deeper MLPs therefore
establish depth/capacity sensitivity rather than a parameter-matched
causal effect of depth. A controlled follow-up should compare deeper
models against width-matched or parameter-matched shallow networks.

The present work also intentionally prioritizes lightweight neural
architectures. This makes the representation trajectory directly
auditable, but does not establish how the same interaction structure
behaves under modern cross-token processors.

A natural next step is to extend ORBIT to Transformer-style
self-attention backbones. Repeated cross-token mixing would allow us to
measure whether higher-order interaction structure is progressively
constructed across attention layers,

\begin{equation}
Z_{\mathrm{tok}}
\rightarrow
H_1
\rightarrow
H_2
\rightarrow
\cdots
\rightarrow
H_L,
\end{equation}

while tracking $E_{\ell,k}$ and $A_{\ell,k}$ at each stage. This would
test whether interaction information is introduced primarily by the
initial token representation, constructed progressively by the
processor, or emerges through their interaction.

The present results motivate this experiment but do not establish that
Transformer processing will necessarily improve higher-order recovery.

Finally, RIT itself is constructed from pairwise interaction functions.
Its significant token-stage pairwise accessibility advantage is
therefore consistent with its design, but the architecture does not
guarantee genuine third- or fourth-order accessibility. Higher-order
structure must be established empirically through ORBIT rather than
inferred from the token representation mechanism.
\section{Conclusion}

We introduce ORBIT, an order-resolved framework for tracing the
emergence of nonlinear interaction structure through learned protein
representations.

Rather than asking only whether a model predicts fitness accurately,
ORBIT asks:

\begin{quote}
\textit{Where do interactions of different orders become accessible,
and does successful prediction imply faithful recovery of the
underlying epistatic structure?}
\end{quote}

By combining exact synthetic systems, biologically motivated GB1
generalization, representation-level Walsh analysis, and
order-specific functional recovery, ORBIT provides a lightweight
framework for studying higher-order molecular representation learning.


\appendix

\section{Reproducibility and Frozen Experimental Protocol}
\label{app:reproducibility}

This appendix documents the frozen experimental and inferential
protocol underlying the results reported in the main text. It records
the data and evaluation artifacts, execution order, stochastic
replication scheme, depth/capacity experiment, probe-fitting procedure,
confirmatory statistical analysis, and provenance controls used for the
final ORBIT experiments. These details are reported separately from the
main methodological description to make the final analysis directly
auditable and reproducible.

\subsection{Data and Evaluation Artifacts}

The GB1 dataset contains 149,361 experimentally measured variants across
the four positions V39, D40, G41, and V54.

The official FLIP 2-vs-rest artifact used for prediction contains

\begin{equation}
424\ \text{training variants}
\qquad\text{and}\qquad
8{,}309\ \text{test variants}.
\end{equation}

The prediction-training partition contains WT, single mutants, and
double mutants ($\mathrm{HD}\leq2$), while the FLIP test partition
contains triple and quadruple mutants ($\mathrm{HD}=3$--$4$).

For ORBIT analysis, all $19^4=130{,}321$ WT-anchored candidate
four-dimensional cubes were enumerated. Requiring all 16 vertices to
have measured fitness retained

\begin{equation}
109{,}235
\end{equation}

complete cubes.

The AA-identity-strict probe construction produced

\begin{equation}
7{,}352\ \text{probe-training cubes},
\end{equation}

\begin{equation}
245\ \text{probe-validation cubes},
\end{equation}

and

\begin{equation}
613\ \text{probe-test cubes}.
\end{equation}

Non-WT amino-acid identities are disjoint across these probe partitions
at each position; the WT identity is intentionally shared.

\subsection{End-to-End Execution Order}

The frozen primary GB1 analysis was executed in the following order:

\begin{enumerate}[leftmargin=*]
    \item validate Walsh-transform unit tests;
    \item validate ORBIT-Synth unit tests;
    \item run the ORBIT-Synth ground-truth benchmark;
    \item prepare the GB1 dataset;
    \item freeze and validate the official FLIP split;
    \item construct complete WT-anchored fitness cubes;
    \item audit cube overlap;
    \item construct the AA-identity-strict probe split;
    \item train the GB1 models;
    \item evaluate FLIP prediction;
    \item compute function-level Walsh recovery;
    \item compute representation accessibility and Walsh energy;
    \item run the pre-specified statistical tests;
    \item generate manuscript tables;
    \item generate manuscript figures;
    \item generate statistical summary tables; and
    \item generate statistical summary figures.
\end{enumerate}

The complete primary pipeline was rerun from clean archived outputs
before final reporting.

\subsection{Stochastic Replication}

All stochastic neural models use the same 20 paired seeds,

\begin{equation}
42,\ldots,61.
\end{equation}

Using paired seeds preserves seed-level correspondence across model and
depth comparisons. Ridge regression is deterministic and is not treated
as a stochastic replicate.

\subsection{Depth/Capacity Experiment}

The depth/capacity experiment evaluates

\begin{equation}
4\ \text{architectures}
\times
3\ \text{depths}
\times
20\ \text{paired seeds}
=
240
\end{equation}

trained neural-model configurations.

The architectures are MLP, Independent tokens, Nonlinear independent
tokens, and RIT. The evaluated backbone depths are 2, 3, and 4.

Depths 3 and 4 add 32-dimensional hidden layers after $H_2$.
Consequently, this analysis tests depth/capacity sensitivity rather than
a parameter-matched causal effect of depth.

The confirmatory depth contrasts were frozen before inspection of the
depth results:

\begin{equation}
d_3-d_2
\qquad\text{and}\qquad
d_4-d_2.
\end{equation}

The $d_4-d_3$ contrast is not treated as confirmatory.

\subsection{Probe Fitting}

For each representation stage and interaction order, ridge-probe
regularization is selected using the probe-validation partition only.
The selected probe is then refit on the union of probe-training and
probe-validation data and evaluated once on the 613 strict probe-test
cubes.

No probe-test results are used for hyperparameter selection.

\subsection{Confirmatory Statistical Inference}

All stochastic confirmatory comparisons use a two-sided exact paired
sign-flip test over

\begin{equation}
2^{20}=1{,}048{,}576
\end{equation}

possible sign assignments.

For each contrast we report the mean paired difference, a 95\% paired
percentile-bootstrap confidence interval based on 20,000 resamples,
paired effect size $d_z$, the raw exact permutation $p$-value, and the
Holm-adjusted $p$-value.

The final frozen analysis contains 57 confirmatory tests grouped into
four multiplicity families:

\begin{itemize}[leftmargin=*]
    \item prediction: 11 tests;
    \item higher-order functional recovery: 22 tests;
    \item final-layer higher-order accessibility: 22 tests;
    \item token-stage pairwise accessibility: 2 tests.
\end{itemize}

Holm correction is performed separately within each family.
Representation Walsh energy and intermediate-layer trajectories are
treated descriptively and do not generate additional confirmatory
hypotheses.

\subsection{Frozen Analysis and Provenance}

The depth hypotheses were frozen before inspection of the depth results.
The archived experimental package contains the frozen hypothesis plan,
analysis scripts, statistical outputs, and machine-readable provenance
manifest with SHA-256 checksums.

This separates pre-specified confirmatory inference from descriptive
post hoc interpretation and provides a direct record of the code and
analysis artifacts used to generate the reported results.

\section{Complete Primary GB1 Results}
\label{app:complete_primary_results}

This appendix provides the complete numerical results for the primary
two-hidden-layer GB1 experiment. The tables report order-resolved
functional recovery, representation accessibility, and normalized
representation Walsh energy for the evaluated architectures and
representation stages.

Functional recovery is reported for interaction orders 1--4, while
representation-level results distinguish the presence of order-specific
variation from its linear accessibility with respect to experimentally
measured Walsh coefficients. For the token-based models, results are
shown at $Z_{\mathrm{tok}}$, $H_1$, and $H_2$; the MLP is evaluated at
$H_1$ and $H_2$. Neural-model entries summarize the same 20 stochastic
training seeds used throughout the primary analysis.

\paragraph{Functional recovery.}
The functional-recovery table shows how faithfully each model reproduces
the experimentally measured Walsh coefficients at each interaction order.
Across the primary models, recovery is strongest at first and second order
and becomes progressively weaker and more variable at third and fourth
order. The nonlinear models substantially outperform the additive Ridge
baseline at the lower orders. At higher orders, descriptive differences
between neural architectures are present, but the pre-specified paired
comparisons do not support a reliable architecture-level advantage.

\begin{table}[H]
\centering
\caption{Order-resolved functional Walsh recovery on the 613 AA-identity-strict probe-test cubes. Values are pooled $R^2$ across coefficients of each interaction order. Neural-model values are mean $\pm$ standard deviation across 20 seeds (42--61); Ridge is deterministic under the frozen configuration.}
\label{tab:gb1_functional_walsh}
\begin{tabular}{lcccc}
\toprule
Model & $R_1^2$ & $R_2^2$ & $R_3^2$ & $R_4^2$ \\
\midrule
Ridge & 0.4126 ± 0.0000 & -0.0012 ± 0.0000 & -0.0147 ± 0.0000 & -0.0218 ± 0.0000 \\
MLP & 0.5397 ± 0.0764 & 0.4969 ± 0.0657 & 0.1230 ± 0.0858 & -0.0423 ± 0.0787 \\
Independent tokens & 0.5498 ± 0.1187 & 0.4903 ± 0.1406 & 0.2117 ± 0.0864 & 0.0190 ± 0.0965 \\
Nonlinear tokens & 0.6356 ± 0.1011 & 0.6769 ± 0.0972 & 0.2257 ± 0.1218 & 0.0129 ± 0.1091 \\
RIT & 0.5516 ± 0.1160 & 0.4951 ± 0.1287 & 0.2043 ± 0.0955 & 0.0545 ± 0.1037 \\
\bottomrule
\end{tabular}
\end{table}

\paragraph{Representation accessibility.}
The accessibility table addresses a different question: whether
experimentally measured order-specific epistasis can be linearly decoded
from each internal representation stage. The most prominent
architecture-specific pattern occurs at the token stage, where RIT makes
pairwise epistatic information substantially more accessible than either
independent-token control. This early advantage does not persist as a
significant third- or fourth-order advantage at the final hidden layer,
illustrating that early accessibility and downstream functional recovery
need not coincide.

\begin{table}[H]
\centering
\caption{Linear accessibility of experimentally derived order-specific Walsh coefficients from learned representations. Probe alpha is selected on probe validation, the probe is refit on probe train plus validation, and evaluated once on the 613 AA-identity-strict probe-test cubes. Values are mean $\pm$ standard deviation across 20 seeds.}
\label{tab:gb1_accessibility}
\begin{tabular}{llcccc}
\toprule
Model & Layer & $A_{\ell,1}$ & $A_{\ell,2}$ & $A_{\ell,3}$ & $A_{\ell,4}$ \\
\midrule
MLP & $H_1$ & 0.7016 ± 0.0348 & 0.4721 ± 0.0675 & 0.0413 ± 0.0808 & -0.3206 ± 0.1733 \\
MLP & $H_2$ & 0.7481 ± 0.0469 & 0.5534 ± 0.0633 & 0.1774 ± 0.0715 & -0.2187 ± 0.2091 \\
Independent tokens & $Z_{\mathrm{token}}$ & 0.3908 ± 0.0798 & -0.0539 ± 0.0000 & -0.0355 ± 0.0000 & -0.1758 ± 0.0000 \\
Independent tokens & $H_1$ & 0.6091 ± 0.0516 & 0.4336 ± 0.0974 & 0.0976 ± 0.1133 & -0.3316 ± 0.1537 \\
Independent tokens & $H_2$ & 0.7491 ± 0.0715 & 0.5791 ± 0.1244 & 0.2245 ± 0.1058 & -0.1726 ± 0.1777 \\
Nonlinear tokens & $Z_{\mathrm{token}}$ & 0.5127 ± 0.0483 & -0.0539 ± 0.0000 & -0.0355 ± 0.0000 & -0.1758 ± 0.0000 \\
Nonlinear tokens & $H_1$ & 0.7889 ± 0.0701 & 0.4094 ± 0.3183 & 0.0817 ± 0.2432 & -0.4549 ± 0.3133 \\
Nonlinear tokens & $H_2$ & 0.7912 ± 0.0789 & 0.6937 ± 0.0953 & 0.2179 ± 0.1176 & -0.2273 ± 0.1319 \\
RIT & $Z_{\mathrm{token}}$ & 0.4302 ± 0.0632 & 0.1929 ± 0.1475 & -0.3334 ± 0.3081 & -0.3468 ± 0.2544 \\
RIT & $H_1$ & 0.6204 ± 0.0573 & 0.4603 ± 0.0747 & 0.1027 ± 0.0784 & -0.3397 ± 0.1993 \\
RIT & $H_2$ & 0.7532 ± 0.0605 & 0.6023 ± 0.1004 & 0.2094 ± 0.1053 & -0.2302 ± 0.2087 \\
\bottomrule
\end{tabular}
\end{table}

\FloatBarrier

\paragraph{Representation Walsh energy.}
The representation-energy table describes where order-specific variation
is present, independently of whether that variation is aligned with the
experimental epistatic coefficients. First-order components dominate most
representations. Independent and Nonlinear-independent tokens contain only
first-order structure at the initial token stage by construction, whereas
RIT introduces second-order energy through its explicit pairwise
interaction mechanism. Subsequent nonlinear layers redistribute some
energy toward higher orders, although third- and fourth-order components
remain comparatively small.

\begin{table}[H]
\centering
\caption{Normalized representation Walsh energy on the 613 strict probe-test cubes. The order-0 component is excluded from the normalization denominator. $E_k$ denotes total energy at order $k$, not average coefficient magnitude. Values are mean $\pm$ standard deviation across 20 seeds.}
\label{tab:gb1_representation_energy}
\begin{tabular}{llcccc}
\toprule
Model & Layer & $E_{\ell,1}$ & $E_{\ell,2}$ & $E_{\ell,3}$ & $E_{\ell,4}$ \\
\midrule
MLP & $H_1$ & 0.937370 $\pm$ 0.006139 & 0.057134 $\pm$ 0.005665 & 0.004840 $\pm$ 0.000519 & 0.000656 $\pm$ 0.000097 \\
MLP & $H_2$ & 0.878389 $\pm$ 0.034318 & 0.111567 $\pm$ 0.029501 & 0.009065 $\pm$ 0.004603 & 0.000979 $\pm$ 0.000480 \\
Independent tokens & $Z_{\mathrm{token}}$ & 1.000000 $\pm$ 0.000000 & 0.000000 $\pm$ 0.000000 & 0.000000 $\pm$ 0.000000 & 0.000000 $\pm$ 0.000000 \\
Independent tokens & $H_1$ & 0.922745 $\pm$ 0.008103 & 0.070863 $\pm$ 0.007632 & 0.005628 $\pm$ 0.000612 & 0.000764 $\pm$ 0.000126 \\
Independent tokens & $H_2$ & 0.856304 $\pm$ 0.038567 & 0.128730 $\pm$ 0.033405 & 0.013499 $\pm$ 0.004856 & 0.001467 $\pm$ 0.000551 \\
Nonlinear tokens & $Z_{\mathrm{token}}$ & 1.000000 $\pm$ 0.000000 & 0.000000 $\pm$ 0.000000 & 0.000000 $\pm$ 0.000000 & 0.000000 $\pm$ 0.000000 \\
Nonlinear tokens & $H_1$ & 0.949630 $\pm$ 0.011800 & 0.046110 $\pm$ 0.011042 & 0.003862 $\pm$ 0.000914 & 0.000397 $\pm$ 0.000122 \\
Nonlinear tokens & $H_2$ & 0.845361 $\pm$ 0.038371 & 0.138808 $\pm$ 0.033379 & 0.014711 $\pm$ 0.005295 & 0.001120 $\pm$ 0.000491 \\
RIT & $Z_{\mathrm{token}}$ & 0.999993 $\pm$ 0.000008 & 0.000007 $\pm$ 0.000008 & 0.000000 $\pm$ 0.000000 & 0.000000 $\pm$ 0.000000 \\
RIT & $H_1$ & 0.918471 $\pm$ 0.011520 & 0.074989 $\pm$ 0.010910 & 0.005776 $\pm$ 0.000549 & 0.000764 $\pm$ 0.000151 \\
RIT & $H_2$ & 0.852641 $\pm$ 0.039291 & 0.132332 $\pm$ 0.034785 & 0.013602 $\pm$ 0.004435 & 0.001425 $\pm$ 0.000441 \\
\bottomrule
\end{tabular}
\end{table}

Taken together, these complementary measurements show why ORBIT does not
collapse representation analysis into a single quantity. Order-specific
variation may be present without being experimentally aligned, and
experimentally aligned information may become accessible at an internal
stage without producing a corresponding improvement in the final
predicted fitness landscape. Complete confirmatory inference for the
primary architecture contrasts is reported in
Appendix~\ref{app:complete_statistics}.
\FloatBarrier

\clearpage
\section{Complete Depth/Capacity Results}
\label{app:depth_results}

This appendix provides the complete numerical results from the
pre-specified depth/capacity sensitivity analysis. The experiment
evaluates MLP, Independent-token, Nonlinear-independent-token, and RIT
models with two, three, and four hidden layers using the same 20 paired
training seeds as the primary analysis.

The tables report four complementary views of the depth experiment:
official FLIP test prediction, order-resolved functional Walsh recovery,
layer- and order-resolved representation accessibility, and normalized
representation Walsh energy. Representation stages are reported through
the final available hidden layer at each depth, allowing the full
trajectory of order-specific structure to be inspected rather than only
the final representation.

Because the three- and four-hidden-layer models contain additional
parameters, these results characterize sensitivity to increased
downstream depth and capacity rather than a parameter-matched causal
effect of depth alone. Representation Walsh energy and intermediate-layer
accessibility trajectories are descriptive; confirmatory inference is
restricted to the pre-specified prediction, higher-order functional
recovery, and final-layer accessibility endpoints defined in the main
text.

Complete statistical results for the corresponding pre-specified depth
contrasts are reported in Appendix~\ref{app:complete_statistics}.

\begin{table}[H]
\centering
\caption{Prediction performance in the ORBIT-GB1 deeper-backbone ablation. Values are mean $\pm$ sample standard deviation across 20 paired training seeds (42--61). Depths 3 and 4 add 32-wide hidden layers, so the comparison measures depth/capacity sensitivity rather than a parameter-matched causal effect of depth.}
\label{tab:gb1_depth_prediction}
\begin{tabular}{llc}
\toprule
Model & Backbone depth & Official FLIP test $R^2$ \\
\midrule
MLP & 2 & 0.1392 $\pm$ 0.1531 \\
MLP & 3 & 0.2729 $\pm$ 0.0729 \\
MLP & 4 & 0.2953 $\pm$ 0.0435 \\
Independent tokens & 2 & 0.2556 $\pm$ 0.0862 \\
Independent tokens & 3 & 0.3110 $\pm$ 0.0520 \\
Independent tokens & 4 & 0.3249 $\pm$ 0.0429 \\
Nonlinear tokens & 2 & 0.2109 $\pm$ 0.1298 \\
Nonlinear tokens & 3 & 0.2338 $\pm$ 0.0743 \\
Nonlinear tokens & 4 & 0.2339 $\pm$ 0.0856 \\
RIT & 2 & 0.2340 $\pm$ 0.1085 \\
RIT & 3 & 0.3004 $\pm$ 0.0534 \\
RIT & 4 & 0.2748 $\pm$ 0.0636 \\
\bottomrule
\end{tabular}
\end{table}

\begin{table}[H]
\centering
\caption{Order-resolved functional Walsh recovery on the 613 AA-identity-strict probe-test cubes in the deeper-backbone ablation. Each value is mean $\pm$ sample standard deviation across 20 paired training seeds (42--61).}
\label{tab:gb1_depth_functional_walsh}
\begin{tabular}{llcccc}
\toprule
Model & Backbone depth & $R_1^2$ & $R_2^2$ & $R_3^2$ & $R_4^2$ \\
\midrule
MLP & 2 & 0.5397 $\pm$ 0.0764 & 0.4969 $\pm$ 0.0657 & 0.1230 $\pm$ 0.0858 & -0.0423 $\pm$ 0.0787 \\
MLP & 3 & 0.7003 $\pm$ 0.0623 & 0.6243 $\pm$ 0.0506 & 0.2963 $\pm$ 0.0727 & -0.0263 $\pm$ 0.0758 \\
MLP & 4 & 0.7081 $\pm$ 0.0492 & 0.6503 $\pm$ 0.0443 & 0.3399 $\pm$ 0.0487 & -0.0042 $\pm$ 0.0872 \\
Independent tokens & 2 & 0.5498 $\pm$ 0.1187 & 0.4903 $\pm$ 0.1406 & 0.2117 $\pm$ 0.0864 & 0.0190 $\pm$ 0.0965 \\
Independent tokens & 3 & 0.6251 $\pm$ 0.1331 & 0.5445 $\pm$ 0.1128 & 0.2493 $\pm$ 0.1074 & -0.0023 $\pm$ 0.1561 \\
Independent tokens & 4 & 0.6589 $\pm$ 0.1099 & 0.5757 $\pm$ 0.1097 & 0.2618 $\pm$ 0.1426 & 0.0064 $\pm$ 0.1675 \\
Nonlinear tokens & 2 & 0.6356 $\pm$ 0.1011 & 0.6769 $\pm$ 0.0972 & 0.2257 $\pm$ 0.1218 & 0.0129 $\pm$ 0.1091 \\
Nonlinear tokens & 3 & 0.6847 $\pm$ 0.0891 & 0.7022 $\pm$ 0.0685 & 0.2406 $\pm$ 0.1176 & 0.0434 $\pm$ 0.0963 \\
Nonlinear tokens & 4 & 0.7175 $\pm$ 0.0769 & 0.7238 $\pm$ 0.0568 & 0.2505 $\pm$ 0.1572 & 0.0145 $\pm$ 0.1138 \\
RIT & 2 & 0.5516 $\pm$ 0.1160 & 0.4951 $\pm$ 0.1287 & 0.2043 $\pm$ 0.0955 & 0.0545 $\pm$ 0.1037 \\
RIT & 3 & 0.6154 $\pm$ 0.0933 & 0.5559 $\pm$ 0.1268 & 0.2458 $\pm$ 0.1115 & -0.0066 $\pm$ 0.1285 \\
RIT & 4 & 0.6036 $\pm$ 0.1686 & 0.5595 $\pm$ 0.1395 & 0.2676 $\pm$ 0.1203 & -0.0452 $\pm$ 0.1420 \\
\bottomrule
\end{tabular}
\end{table}

\begin{table}[H]
\centering
\caption{Layer- and order-resolved linear accessibility in the deeper-backbone ablation. The same AA-identity-strict probe protocol is used at every stage. Values are mean $\pm$ sample standard deviation across 20 paired training seeds (42--61).}
\label{tab:gb1_depth_accessibility}
\begin{tabular}{lllcccc}
\toprule
Model & Depth & Stage & $A_{\ell,1}$ & $A_{\ell,2}$ & $A_{\ell,3}$ & $A_{\ell,4}$ \\
\midrule
MLP & 2 & $H_1$ & 0.7016 $\pm$ 0.0348 & 0.4721 $\pm$ 0.0675 & 0.0413 $\pm$ 0.0808 & -0.3206 $\pm$ 0.1733 \\
MLP & 2 & $H_2$ & 0.7481 $\pm$ 0.0469 & 0.5534 $\pm$ 0.0633 & 0.1774 $\pm$ 0.0715 & -0.2187 $\pm$ 0.2091 \\
MLP & 3 & $H_1$ & 0.6383 $\pm$ 0.0531 & 0.4138 $\pm$ 0.1491 & 0.0047 $\pm$ 0.1240 & -0.4583 $\pm$ 0.5004 \\
MLP & 3 & $H_2$ & 0.7632 $\pm$ 0.0453 & 0.5663 $\pm$ 0.0737 & 0.1977 $\pm$ 0.0762 & -0.1943 $\pm$ 0.1115 \\
MLP & 3 & $H_3$ & 0.7998 $\pm$ 0.0337 & 0.6669 $\pm$ 0.0408 & 0.2831 $\pm$ 0.0604 & -0.0535 $\pm$ 0.0639 \\
MLP & 4 & $H_1$ & 0.6419 $\pm$ 0.0559 & 0.4104 $\pm$ 0.0622 & 0.0193 $\pm$ 0.0636 & -0.2850 $\pm$ 0.1378 \\
MLP & 4 & $H_2$ & 0.7688 $\pm$ 0.0445 & 0.5656 $\pm$ 0.1068 & 0.1578 $\pm$ 0.0887 & -0.1963 $\pm$ 0.0872 \\
MLP & 4 & $H_3$ & 0.7953 $\pm$ 0.0347 & 0.6654 $\pm$ 0.0476 & 0.3048 $\pm$ 0.0494 & -0.0960 $\pm$ 0.1234 \\
MLP & 4 & $H_4$ & 0.7925 $\pm$ 0.0442 & 0.6705 $\pm$ 0.0347 & 0.3074 $\pm$ 0.0581 & -0.0417 $\pm$ 0.0946 \\
Independent tokens & 2 & $Z_{\mathrm{token}}$ & 0.3908 $\pm$ 0.0798 & -0.0539 $\pm$ 0.0000 & -0.0355 $\pm$ 0.0000 & -0.1758 $\pm$ 0.0000 \\
Independent tokens & 2 & $H_1$ & 0.6091 $\pm$ 0.0516 & 0.4336 $\pm$ 0.0974 & 0.0976 $\pm$ 0.1133 & -0.3316 $\pm$ 0.1537 \\
Independent tokens & 2 & $H_2$ & 0.7491 $\pm$ 0.0715 & 0.5791 $\pm$ 0.1244 & 0.2245 $\pm$ 0.1058 & -0.1726 $\pm$ 0.1777 \\
Independent tokens & 3 & $Z_{\mathrm{token}}$ & 0.3853 $\pm$ 0.0813 & -0.0539 $\pm$ 0.0000 & -0.0355 $\pm$ 0.0000 & -0.1758 $\pm$ 0.0000 \\
Independent tokens & 3 & $H_1$ & 0.5794 $\pm$ 0.0870 & 0.3909 $\pm$ 0.0738 & 0.0503 $\pm$ 0.0976 & -0.3608 $\pm$ 0.1840 \\
Independent tokens & 3 & $H_2$ & 0.7368 $\pm$ 0.0915 & 0.5289 $\pm$ 0.2099 & 0.1457 $\pm$ 0.2096 & -0.2312 $\pm$ 0.2511 \\
Independent tokens & 3 & $H_3$ & 0.7630 $\pm$ 0.0757 & 0.6037 $\pm$ 0.0934 & 0.2351 $\pm$ 0.1137 & -0.1095 $\pm$ 0.1301 \\
Independent tokens & 4 & $Z_{\mathrm{token}}$ & 0.3810 $\pm$ 0.0830 & -0.0539 $\pm$ 0.0000 & -0.0355 $\pm$ 0.0000 & -0.1758 $\pm$ 0.0000 \\
Independent tokens & 4 & $H_1$ & 0.5698 $\pm$ 0.0689 & 0.3687 $\pm$ 0.0730 & 0.0656 $\pm$ 0.1128 & -0.3166 $\pm$ 0.1874 \\
Independent tokens & 4 & $H_2$ & 0.7377 $\pm$ 0.0749 & 0.5414 $\pm$ 0.1065 & 0.1462 $\pm$ 0.1290 & -0.2377 $\pm$ 0.1928 \\
Independent tokens & 4 & $H_3$ & 0.7953 $\pm$ 0.0597 & 0.6239 $\pm$ 0.0885 & 0.2298 $\pm$ 0.1296 & -0.1266 $\pm$ 0.1994 \\
Independent tokens & 4 & $H_4$ & 0.8070 $\pm$ 0.0621 & 0.6463 $\pm$ 0.0886 & 0.2522 $\pm$ 0.1312 & -0.0675 $\pm$ 0.1515 \\
Nonlinear tokens & 2 & $Z_{\mathrm{token}}$ & 0.5127 $\pm$ 0.0483 & -0.0539 $\pm$ 0.0000 & -0.0355 $\pm$ 0.0000 & -0.1758 $\pm$ 0.0000 \\
Nonlinear tokens & 2 & $H_1$ & 0.7889 $\pm$ 0.0701 & 0.4094 $\pm$ 0.3183 & 0.0817 $\pm$ 0.2432 & -0.4549 $\pm$ 0.3133 \\
Nonlinear tokens & 2 & $H_2$ & 0.7912 $\pm$ 0.0789 & 0.6937 $\pm$ 0.0953 & 0.2179 $\pm$ 0.1176 & -0.2273 $\pm$ 0.1319 \\
Nonlinear tokens & 3 & $Z_{\mathrm{token}}$ & 0.4961 $\pm$ 0.0631 & -0.0539 $\pm$ 0.0000 & -0.0355 $\pm$ 0.0000 & -0.1758 $\pm$ 0.0000 \\
Nonlinear tokens & 3 & $H_1$ & 0.7734 $\pm$ 0.0809 & 0.2947 $\pm$ 0.9017 & 0.0416 $\pm$ 0.2699 & -0.6515 $\pm$ 1.3515 \\
Nonlinear tokens & 3 & $H_2$ & 0.8186 $\pm$ 0.0366 & 0.7059 $\pm$ 0.0631 & 0.2180 $\pm$ 0.1110 & -0.1956 $\pm$ 0.1231 \\
Nonlinear tokens & 3 & $H_3$ & 0.8142 $\pm$ 0.0561 & 0.7082 $\pm$ 0.0651 & 0.2399 $\pm$ 0.1130 & -0.1964 $\pm$ 0.1076 \\
Nonlinear tokens & 4 & $Z_{\mathrm{token}}$ & -8.0518 $\pm$ 37.1861 & -0.0539 $\pm$ 0.0000 & -0.0355 $\pm$ 0.0000 & -0.1758 $\pm$ 0.0000 \\
Nonlinear tokens & 4 & $H_1$ & 0.7719 $\pm$ 0.0607 & 0.4116 $\pm$ 0.3245 & -0.1555 $\pm$ 0.6327 & -1.5320 $\pm$ 3.3408 \\
Nonlinear tokens & 4 & $H_2$ & 0.7355 $\pm$ 0.4067 & -0.6521 $\pm$ 6.1046 & 0.2018 $\pm$ 0.2135 & -1.2509 $\pm$ 4.6885 \\
Nonlinear tokens & 4 & $H_3$ & 0.8218 $\pm$ 0.0536 & 0.7357 $\pm$ 0.0517 & 0.2898 $\pm$ 0.1156 & -0.2446 $\pm$ 0.1843 \\
Nonlinear tokens & 4 & $H_4$ & 0.8034 $\pm$ 0.0603 & 0.7178 $\pm$ 0.0571 & 0.2637 $\pm$ 0.1295 & -0.1943 $\pm$ 0.1282 \\
RIT & 2 & $Z_{\mathrm{token}}$ & 0.4302 $\pm$ 0.0632 & 0.1929 $\pm$ 0.1475 & -0.3334 $\pm$ 0.3081 & -0.3468 $\pm$ 0.2544 \\
RIT & 2 & $H_1$ & 0.6204 $\pm$ 0.0573 & 0.4603 $\pm$ 0.0747 & 0.1027 $\pm$ 0.0784 & -0.3397 $\pm$ 0.1993 \\
RIT & 2 & $H_2$ & 0.7532 $\pm$ 0.0605 & 0.6023 $\pm$ 0.1004 & 0.2094 $\pm$ 0.1053 & -0.2302 $\pm$ 0.2087 \\
RIT & 3 & $Z_{\mathrm{token}}$ & 0.4020 $\pm$ 0.0695 & 0.1531 $\pm$ 0.1527 & -0.4135 $\pm$ 0.5199 & -0.4051 $\pm$ 0.3181 \\
RIT & 3 & $H_1$ & 0.5889 $\pm$ 0.0923 & 0.4274 $\pm$ 0.0706 & 0.0672 $\pm$ 0.0893 & -0.3250 $\pm$ 0.1452 \\
RIT & 3 & $H_2$ & 0.7329 $\pm$ 0.0727 & 0.5839 $\pm$ 0.0898 & -0.0525 $\pm$ 0.9347 & -1.3024 $\pm$ 4.8170 \\
RIT & 3 & $H_3$ & 0.7914 $\pm$ 0.0485 & 0.6426 $\pm$ 0.0851 & 0.2412 $\pm$ 0.0739 & -0.1003 $\pm$ 0.1316 \\
RIT & 4 & $Z_{\mathrm{token}}$ & 0.3950 $\pm$ 0.0794 & 0.0903 $\pm$ 0.2133 & -0.2287 $\pm$ 0.3339 & -0.3211 $\pm$ 0.2802 \\
RIT & 4 & $H_1$ & 0.6002 $\pm$ 0.0551 & 0.4080 $\pm$ 0.0625 & 0.0655 $\pm$ 0.0884 & -0.3379 $\pm$ 0.1435 \\
RIT & 4 & $H_2$ & 0.7340 $\pm$ 0.0532 & 0.5530 $\pm$ 0.0893 & 0.1876 $\pm$ 0.1013 & -0.2346 $\pm$ 0.1370 \\
RIT & 4 & $H_3$ & 0.7749 $\pm$ 0.0649 & 0.6289 $\pm$ 0.0979 & 0.2279 $\pm$ 0.1203 & -0.0763 $\pm$ 0.1322 \\
RIT & 4 & $H_4$ & 0.7892 $\pm$ 0.0697 & 0.6374 $\pm$ 0.0957 & 0.2275 $\pm$ 0.1713 & -0.0469 $\pm$ 0.1655 \\
\bottomrule
\end{tabular}
\end{table}

\begin{table}[H]
\centering
\caption{Normalized representation Walsh energy in the deeper-backbone ablation. The order-0 component is excluded from the normalization denominator; $E_{\ell,k}$ is a descriptive representation-energy metric, not a biochemical interaction-strength estimate. Values are mean $\pm$ sample standard deviation across 20 paired training seeds (42--61).}
\label{tab:gb1_depth_representation_energy}
\resizebox{\textwidth}{!}{%
\begin{tabular}{lllcccc}
\toprule
Model & Depth & Stage & $E_{\ell,1}$ & $E_{\ell,2}$ & $E_{\ell,3}$ & $E_{\ell,4}$ \\
\midrule
MLP & 2 & $H_1$ & 0.937370 $\pm$ 0.006139 & 0.057134 $\pm$ 0.005665 & 0.004840 $\pm$ 0.000519 & 0.000656 $\pm$ 0.000097 \\
MLP & 2 & $H_2$ & 0.878389 $\pm$ 0.034318 & 0.111567 $\pm$ 0.029501 & 0.009065 $\pm$ 0.004603 & 0.000979 $\pm$ 0.000480 \\
MLP & 3 & $H_1$ & 0.944213 $\pm$ 0.004841 & 0.050727 $\pm$ 0.004442 & 0.004466 $\pm$ 0.000424 & 0.000595 $\pm$ 0.000095 \\
MLP & 3 & $H_2$ & 0.918602 $\pm$ 0.019442 & 0.076230 $\pm$ 0.018444 & 0.004638 $\pm$ 0.001154 & 0.000530 $\pm$ 0.000154 \\
MLP & 3 & $H_3$ & 0.806304 $\pm$ 0.025654 & 0.169747 $\pm$ 0.021498 & 0.021968 $\pm$ 0.004587 & 0.001981 $\pm$ 0.000457 \\
MLP & 4 & $H_1$ & 0.945460 $\pm$ 0.004794 & 0.049532 $\pm$ 0.004442 & 0.004416 $\pm$ 0.000406 & 0.000593 $\pm$ 0.000079 \\
MLP & 4 & $H_2$ & 0.932448 $\pm$ 0.016934 & 0.062848 $\pm$ 0.015257 & 0.004239 $\pm$ 0.001853 & 0.000465 $\pm$ 0.000159 \\
MLP & 4 & $H_3$ & 0.874361 $\pm$ 0.030773 & 0.114294 $\pm$ 0.026456 & 0.010440 $\pm$ 0.004359 & 0.000904 $\pm$ 0.000308 \\
MLP & 4 & $H_4$ & 0.790237 $\pm$ 0.032651 & 0.179857 $\pm$ 0.026453 & 0.027734 $\pm$ 0.006754 & 0.002173 $\pm$ 0.000581 \\
Independent tokens & 2 & $Z_{\mathrm{token}}$ & 1.000000 $\pm$ 0.000000 & 0.000000 $\pm$ 0.000000 & 0.000000 $\pm$ 0.000000 & 0.000000 $\pm$ 0.000000 \\
Independent tokens & 2 & $H_1$ & 0.922745 $\pm$ 0.008103 & 0.070863 $\pm$ 0.007632 & 0.005628 $\pm$ 0.000612 & 0.000764 $\pm$ 0.000126 \\
Independent tokens & 2 & $H_2$ & 0.856304 $\pm$ 0.038567 & 0.128730 $\pm$ 0.033405 & 0.013499 $\pm$ 0.004856 & 0.001467 $\pm$ 0.000551 \\
Independent tokens & 3 & $Z_{\mathrm{token}}$ & 1.000000 $\pm$ 0.000000 & 0.000000 $\pm$ 0.000000 & 0.000000 $\pm$ 0.000000 & 0.000000 $\pm$ 0.000000 \\
Independent tokens & 3 & $H_1$ & 0.932010 $\pm$ 0.008604 & 0.062169 $\pm$ 0.008013 & 0.005146 $\pm$ 0.000603 & 0.000674 $\pm$ 0.000132 \\
Independent tokens & 3 & $H_2$ & 0.909889 $\pm$ 0.024845 & 0.081982 $\pm$ 0.022571 & 0.007286 $\pm$ 0.002206 & 0.000842 $\pm$ 0.000303 \\
Independent tokens & 3 & $H_3$ & 0.810329 $\pm$ 0.048222 & 0.169700 $\pm$ 0.040859 & 0.018188 $\pm$ 0.007719 & 0.001783 $\pm$ 0.000873 \\
Independent tokens & 4 & $Z_{\mathrm{token}}$ & 1.000000 $\pm$ 0.000000 & 0.000000 $\pm$ 0.000000 & 0.000000 $\pm$ 0.000000 & 0.000000 $\pm$ 0.000000 \\
Independent tokens & 4 & $H_1$ & 0.938384 $\pm$ 0.008950 & 0.056152 $\pm$ 0.008384 & 0.004846 $\pm$ 0.000642 & 0.000619 $\pm$ 0.000093 \\
Independent tokens & 4 & $H_2$ & 0.928759 $\pm$ 0.019462 & 0.065141 $\pm$ 0.018108 & 0.005460 $\pm$ 0.001456 & 0.000641 $\pm$ 0.000172 \\
Independent tokens & 4 & $H_3$ & 0.869924 $\pm$ 0.042093 & 0.119620 $\pm$ 0.039262 & 0.009488 $\pm$ 0.003554 & 0.000968 $\pm$ 0.000315 \\
Independent tokens & 4 & $H_4$ & 0.773630 $\pm$ 0.052995 & 0.200181 $\pm$ 0.045783 & 0.024109 $\pm$ 0.008176 & 0.002080 $\pm$ 0.000749 \\
Nonlinear tokens & 2 & $Z_{\mathrm{token}}$ & 1.000000 $\pm$ 0.000000 & 0.000000 $\pm$ 0.000000 & 0.000000 $\pm$ 0.000000 & 0.000000 $\pm$ 0.000000 \\
Nonlinear tokens & 2 & $H_1$ & 0.949630 $\pm$ 0.011800 & 0.046110 $\pm$ 0.011042 & 0.003862 $\pm$ 0.000914 & 0.000397 $\pm$ 0.000122 \\
Nonlinear tokens & 2 & $H_2$ & 0.845361 $\pm$ 0.038371 & 0.138808 $\pm$ 0.033379 & 0.014711 $\pm$ 0.005295 & 0.001120 $\pm$ 0.000491 \\
Nonlinear tokens & 3 & $Z_{\mathrm{token}}$ & 1.000000 $\pm$ 0.000000 & 0.000000 $\pm$ 0.000000 & 0.000000 $\pm$ 0.000000 & 0.000000 $\pm$ 0.000000 \\
Nonlinear tokens & 3 & $H_1$ & 0.963286 $\pm$ 0.011526 & 0.033704 $\pm$ 0.010886 & 0.002748 $\pm$ 0.000774 & 0.000261 $\pm$ 0.000084 \\
Nonlinear tokens & 3 & $H_2$ & 0.923384 $\pm$ 0.030029 & 0.071929 $\pm$ 0.028808 & 0.004345 $\pm$ 0.001690 & 0.000342 $\pm$ 0.000192 \\
Nonlinear tokens & 3 & $H_3$ & 0.833816 $\pm$ 0.048796 & 0.144605 $\pm$ 0.041788 & 0.020034 $\pm$ 0.007770 & 0.001545 $\pm$ 0.000662 \\
Nonlinear tokens & 4 & $Z_{\mathrm{token}}$ & 1.000000 $\pm$ 0.000000 & 0.000000 $\pm$ 0.000000 & 0.000000 $\pm$ 0.000000 & 0.000000 $\pm$ 0.000000 \\
Nonlinear tokens & 4 & $H_1$ & 0.964249 $\pm$ 0.010711 & 0.033036 $\pm$ 0.010013 & 0.002435 $\pm$ 0.000714 & 0.000280 $\pm$ 0.000233 \\
Nonlinear tokens & 4 & $H_2$ & 0.921552 $\pm$ 0.027599 & 0.074431 $\pm$ 0.026422 & 0.003628 $\pm$ 0.001533 & 0.000389 $\pm$ 0.000302 \\
Nonlinear tokens & 4 & $H_3$ & 0.861184 $\pm$ 0.037826 & 0.127568 $\pm$ 0.033528 & 0.010404 $\pm$ 0.004543 & 0.000845 $\pm$ 0.000528 \\
Nonlinear tokens & 4 & $H_4$ & 0.791632 $\pm$ 0.042018 & 0.179425 $\pm$ 0.035648 & 0.026795 $\pm$ 0.007445 & 0.002148 $\pm$ 0.000934 \\
RIT & 2 & $Z_{\mathrm{token}}$ & 0.999993 $\pm$ 0.000008 & 0.000007 $\pm$ 0.000008 & 0.000000 $\pm$ 0.000000 & 0.000000 $\pm$ 0.000000 \\
RIT & 2 & $H_1$ & 0.918471 $\pm$ 0.011520 & 0.074989 $\pm$ 0.010910 & 0.005776 $\pm$ 0.000549 & 0.000764 $\pm$ 0.000151 \\
RIT & 2 & $H_2$ & 0.852641 $\pm$ 0.039291 & 0.132332 $\pm$ 0.034785 & 0.013602 $\pm$ 0.004435 & 0.001425 $\pm$ 0.000441 \\
RIT & 3 & $Z_{\mathrm{token}}$ & 0.999999 $\pm$ 0.000001 & 0.000001 $\pm$ 0.000001 & 0.000000 $\pm$ 0.000000 & 0.000000 $\pm$ 0.000000 \\
RIT & 3 & $H_1$ & 0.927286 $\pm$ 0.009729 & 0.066650 $\pm$ 0.009155 & 0.005358 $\pm$ 0.000623 & 0.000707 $\pm$ 0.000111 \\
RIT & 3 & $H_2$ & 0.897444 $\pm$ 0.026828 & 0.093739 $\pm$ 0.024782 & 0.007942 $\pm$ 0.002173 & 0.000876 $\pm$ 0.000226 \\
RIT & 3 & $H_3$ & 0.798484 $\pm$ 0.052272 & 0.179259 $\pm$ 0.044776 & 0.020428 $\pm$ 0.008180 & 0.001829 $\pm$ 0.000585 \\
RIT & 4 & $Z_{\mathrm{token}}$ & 1.000000 $\pm$ 0.000000 & 0.000000 $\pm$ 0.000000 & 0.000000 $\pm$ 0.000000 & 0.000000 $\pm$ 0.000000 \\
RIT & 4 & $H_1$ & 0.935997 $\pm$ 0.011001 & 0.058431 $\pm$ 0.010144 & 0.004923 $\pm$ 0.000780 & 0.000649 $\pm$ 0.000154 \\
RIT & 4 & $H_2$ & 0.921058 $\pm$ 0.024282 & 0.072660 $\pm$ 0.022447 & 0.005650 $\pm$ 0.001804 & 0.000632 $\pm$ 0.000213 \\
RIT & 4 & $H_3$ & 0.859682 $\pm$ 0.045986 & 0.128283 $\pm$ 0.041055 & 0.011041 $\pm$ 0.005361 & 0.000994 $\pm$ 0.000412 \\
RIT & 4 & $H_4$ & 0.757734 $\pm$ 0.052513 & 0.212853 $\pm$ 0.044375 & 0.027211 $\pm$ 0.010056 & 0.002202 $\pm$ 0.000842 \\
\bottomrule
\end{tabular}
}
\end{table}

\FloatBarrier

\clearpage
\section{Complete Confirmatory Statistical Results}
\label{app:complete_statistics}

This appendix reports the complete pre-specified confirmatory
statistical analysis underlying the primary architecture comparison
and the depth/capacity sensitivity experiment.

The final frozen hypothesis set contains 57 seed-paired comparisons.
For each comparison, we report the mean paired difference, the
95\% paired-bootstrap confidence interval, paired standardized effect
size $d_z$, the raw two-sided exact paired sign-flip permutation
$p$-value, and the Holm-adjusted $p$-value.

Holm correction is performed separately within the four
pre-specified multiplicity families:

\begin{itemize}[leftmargin=*]
    \item prediction: 11 tests;
    \item higher-order functional recovery: 22 tests;
    \item final-layer higher-order accessibility: 22 tests; and
    \item token-stage pairwise accessibility: 2 tests.
\end{itemize}

The architecture-level comparisons correspond to the frozen primary
two-hidden-layer experiment. The depth/capacity comparisons evaluate,
within each neural architecture, the pre-specified contrasts
depth 3 minus depth 2 and depth 4 minus depth 2. Depth 4 minus
depth 3 was not designated as a confirmatory comparison.

Representation Walsh energy and intermediate-layer accessibility
trajectories are descriptive and are therefore excluded from the
confirmatory hypothesis families. All pre-specified comparisons are
reported below regardless of statistical significance, effect
direction, or whether additional depth improved the corresponding
endpoint.


\begingroup
\footnotesize

\setlength{\tabcolsep}{2.5pt}
\renewcommand{\arraystretch}{1.08}

\setlength{\LTleft}{0pt}
\setlength{\LTright}{0pt}
\setlength{\LTcapwidth}{\linewidth}

\begin{longtable}{@{}
p{0.18\linewidth}
p{0.20\linewidth}
p{0.07\linewidth}
p{0.16\linewidth}
p{0.07\linewidth}
p{0.11\linewidth}
p{0.11\linewidth}
@{}}
\caption{Complete pre-specified confirmatory statistical results. Differences are defined as Model A $-$ Model B for architecture contrasts and depth A $-$ depth B for depth contrasts. Confidence intervals are paired-bootstrap 95\% intervals; $d_z$ is the standardized paired effect size. $p_{\mathrm{raw}}$ is the two-sided exact paired sign-flip permutation $p$-value and $p_{\mathrm{Holm}}$ is adjusted within the corresponding pre-specified family. Bold adjusted $p$-values satisfy $p_{\mathrm{Holm}}\leq0.05$.}
\label{tab:full_confirmatory_stats}\\
\toprule
Endpoint & Comparison & $\Delta$ & 95\% CI & $d_z$ & $p_{\mathrm{raw}}$ & $p_{\mathrm{Holm}}$ \\
\midrule
\endfirsthead
\multicolumn{7}{l}{\textit{Table \thetable\ continued}}\\
\toprule
Endpoint & Comparison & $\Delta$ & 95\% CI & $d_z$ & $p_{\mathrm{raw}}$ & $p_{\mathrm{Holm}}$ \\
\midrule
\endhead
\midrule
\multicolumn{7}{r}{\textit{Continued on next page}}\\
\endfoot
\bottomrule
\endlastfoot
\multicolumn{7}{@{}l}{\textbf{Prediction (11 tests)}}\\
\addlinespace[2pt]
FLIP test $R^2$ & Nonlinear $-$ Independent & -0.045 & [-0.106, 0.014] & -0.320 & 0.172 & 1.000 \\
FLIP test $R^2$ & RIT $-$ Independent & -0.022 & [-0.066, 0.019] & -0.222 & 0.344 & 1.000 \\
FLIP test $R^2$ & RIT $-$ Nonlinear & 0.023 & [-0.049, 0.097] & 0.136 & 0.548 & 1.000 \\
FLIP test $R^2$ & MLP: $d_{3}-d_{2}$ & 0.134 & [0.071, 0.201] & 0.876 & $5.68\times 10^{-4}$ & \textbf{0.006} \\
FLIP test $R^2$ & MLP: $d_{4}-d_{2}$ & 0.156 & [0.097, 0.221] & 1.076 & $6.1\times 10^{-5}$ & \textbf{$6.71\times 10^{-4}$} \\
FLIP test $R^2$ & Independent: $d_{3}-d_{2}$ & 0.055 & [0.015, 0.097] & 0.581 & 0.015 & 0.105 \\
FLIP test $R^2$ & Independent: $d_{4}-d_{2}$ & 0.069 & [0.034, 0.106] & 0.816 & 0.002 & \textbf{0.015} \\
FLIP test $R^2$ & Nonlinear: $d_{3}-d_{2}$ & 0.023 & [-0.022, 0.068] & 0.215 & 0.348 & 1.000 \\
FLIP test $R^2$ & Nonlinear: $d_{4}-d_{2}$ & 0.023 & [-0.026, 0.077] & 0.191 & 0.418 & 1.000 \\
FLIP test $R^2$ & RIT: $d_{3}-d_{2}$ & 0.066 & [0.021, 0.116] & 0.598 & 0.010 & 0.081 \\
FLIP test $R^2$ & RIT: $d_{4}-d_{2}$ & 0.041 & [-0.009, 0.098] & 0.323 & 0.171 & 1.000 \\
\addlinespace[4pt]
\multicolumn{7}{@{}l}{\textbf{Higher-order functional recovery (22 tests)}}\\
\addlinespace[2pt]
Functional $R_3^2$ & Nonlinear $-$ Independent & 0.014 & [-0.051, 0.077] & 0.092 & 0.685 & 1.000 \\
Functional $R_3^2$ & RIT $-$ Independent & -0.007 & [-0.046, 0.031] & -0.082 & 0.717 & 1.000 \\
Functional $R_3^2$ & RIT $-$ Nonlinear & -0.021 & [-0.079, 0.038] & -0.156 & 0.491 & 1.000 \\
Functional $R_4^2$ & Nonlinear $-$ Independent & -0.006 & [-0.064, 0.054] & -0.044 & 0.848 & 1.000 \\
Functional $R_4^2$ & RIT $-$ Independent & 0.036 & [-0.007, 0.077] & 0.360 & 0.124 & 1.000 \\
Functional $R_4^2$ & RIT $-$ Nonlinear & 0.042 & [-0.019, 0.101] & 0.301 & 0.197 & 1.000 \\
Functional $R_3^2$ & MLP: $d_{3}-d_{2}$ & 0.173 & [0.138, 0.207] & 2.115 & $1.91\times 10^{-6}$ & \textbf{$4.2\times 10^{-5}$} \\
Functional $R_3^2$ & MLP: $d_{4}-d_{2}$ & 0.217 & [0.180, 0.256] & 2.404 & $1.91\times 10^{-6}$ & \textbf{$4.2\times 10^{-5}$} \\
Functional $R_3^2$ & Independent: $d_{3}-d_{2}$ & 0.038 & [0.007, 0.069] & 0.520 & 0.032 & 0.601 \\
Functional $R_3^2$ & Independent: $d_{4}-d_{2}$ & 0.050 & [0.000, 0.098] & 0.440 & 0.065 & 1.000 \\
Functional $R_3^2$ & Nonlinear: $d_{3}-d_{2}$ & 0.015 & [-0.031, 0.056] & 0.148 & 0.521 & 1.000 \\
Functional $R_3^2$ & Nonlinear: $d_{4}-d_{2}$ & 0.025 & [-0.030, 0.079] & 0.193 & 0.401 & 1.000 \\
Functional $R_3^2$ & RIT: $d_{3}-d_{2}$ & 0.042 & [-0.011, 0.092] & 0.350 & 0.134 & 1.000 \\
Functional $R_3^2$ & RIT: $d_{4}-d_{2}$ & 0.063 & [0.004, 0.122] & 0.457 & 0.056 & 0.999 \\
Functional $R_4^2$ & MLP: $d_{3}-d_{2}$ & 0.016 & [-0.026, 0.061] & 0.154 & 0.500 & 1.000 \\
Functional $R_4^2$ & MLP: $d_{4}-d_{2}$ & 0.038 & [-0.001, 0.077] & 0.418 & 0.078 & 1.000 \\
Functional $R_4^2$ & Independent: $d_{3}-d_{2}$ & -0.021 & [-0.098, 0.047] & -0.126 & 0.590 & 1.000 \\
Functional $R_4^2$ & Independent: $d_{4}-d_{2}$ & -0.013 & [-0.094, 0.072] & -0.065 & 0.772 & 1.000 \\
Functional $R_4^2$ & Nonlinear: $d_{3}-d_{2}$ & 0.031 & [-0.007, 0.071] & 0.330 & 0.158 & 1.000 \\
Functional $R_4^2$ & Nonlinear: $d_{4}-d_{2}$ & 0.002 & [-0.067, 0.062] & 0.011 & 0.964 & 1.000 \\
Functional $R_4^2$ & RIT: $d_{3}-d_{2}$ & -0.061 & [-0.135, 0.010] & -0.360 & 0.124 & 1.000 \\
Functional $R_4^2$ & RIT: $d_{4}-d_{2}$ & -0.100 & [-0.169, -0.034] & -0.630 & 0.010 & 0.196 \\
\addlinespace[4pt]
\multicolumn{7}{@{}l}{\textbf{Final-layer higher-order accessibility (22 tests)}}\\
\addlinespace[2pt]
$H_2$ accessibility $A_3$ & Nonlinear $-$ Independent & -0.007 & [-0.065, 0.050] & -0.049 & 0.828 & 1.000 \\
$H_2$ accessibility $A_3$ & RIT $-$ Independent & -0.015 & [-0.055, 0.022] & -0.168 & 0.461 & 1.000 \\
$H_2$ accessibility $A_3$ & RIT $-$ Nonlinear & -0.009 & [-0.063, 0.050] & -0.064 & 0.779 & 1.000 \\
$H_2$ accessibility $A_4$ & Nonlinear $-$ Independent & -0.055 & [-0.156, 0.046] & -0.231 & 0.317 & 1.000 \\
$H_2$ accessibility $A_4$ & RIT $-$ Independent & -0.058 & [-0.158, 0.046] & -0.240 & 0.300 & 1.000 \\
$H_2$ accessibility $A_4$ & RIT $-$ Nonlinear & -0.003 & [-0.125, 0.119] & -0.010 & 0.964 & 1.000 \\
Final-layer $A_3$ & MLP: $d_{3}-d_{2}$ & 0.106 & [0.070, 0.138] & 1.323 & $7.06\times 10^{-5}$ & \textbf{0.001} \\
Final-layer $A_3$ & MLP: $d_{4}-d_{2}$ & 0.130 & [0.087, 0.173] & 1.280 & $3.24\times 10^{-5}$ & \textbf{$7.13\times 10^{-4}$} \\
Final-layer $A_3$ & Independent: $d_{3}-d_{2}$ & 0.011 & [-0.023, 0.043] & 0.137 & 0.547 & 1.000 \\
Final-layer $A_3$ & Independent: $d_{4}-d_{2}$ & 0.028 & [-0.021, 0.076] & 0.245 & 0.288 & 1.000 \\
Final-layer $A_3$ & Nonlinear: $d_{3}-d_{2}$ & 0.022 & [-0.032, 0.070] & 0.188 & 0.423 & 1.000 \\
Final-layer $A_3$ & Nonlinear: $d_{4}-d_{2}$ & 0.046 & [-0.011, 0.106] & 0.335 & 0.151 & 1.000 \\
Final-layer $A_3$ & RIT: $d_{3}-d_{2}$ & 0.032 & [-0.009, 0.073] & 0.333 & 0.156 & 1.000 \\
Final-layer $A_3$ & RIT: $d_{4}-d_{2}$ & 0.018 & [-0.069, 0.093] & 0.096 & 0.698 & 1.000 \\
Final-layer $A_4$ & MLP: $d_{3}-d_{2}$ & 0.165 & [0.084, 0.262] & 0.797 & $5.05\times 10^{-4}$ & \textbf{0.010} \\
Final-layer $A_4$ & MLP: $d_{4}-d_{2}$ & 0.177 & [0.087, 0.289] & 0.748 & $9.73\times 10^{-5}$ & \textbf{0.002} \\
Final-layer $A_4$ & Independent: $d_{3}-d_{2}$ & 0.063 & [-0.012, 0.138] & 0.353 & 0.133 & 1.000 \\
Final-layer $A_4$ & Independent: $d_{4}-d_{2}$ & 0.105 & [0.029, 0.181] & 0.592 & 0.016 & 0.252 \\
Final-layer $A_4$ & Nonlinear: $d_{3}-d_{2}$ & 0.031 & [-0.032, 0.093] & 0.212 & 0.356 & 1.000 \\
Final-layer $A_4$ & Nonlinear: $d_{4}-d_{2}$ & 0.033 & [-0.034, 0.103] & 0.205 & 0.373 & 1.000 \\
Final-layer $A_4$ & RIT: $d_{3}-d_{2}$ & 0.130 & [0.046, 0.227] & 0.610 & 0.007 & 0.111 \\
Final-layer $A_4$ & RIT: $d_{4}-d_{2}$ & 0.183 & [0.077, 0.290] & 0.737 & 0.004 & 0.071 \\
\addlinespace[4pt]
\multicolumn{7}{@{}l}{\textbf{Token-stage pairwise accessibility (2 tests)}}\\
\addlinespace[2pt]
$Z_{\mathrm{tok}}$ accessibility $A_2$ & RIT $-$ Independent & 0.247 & [0.183, 0.309] & 1.673 & $5.72\times 10^{-6}$ & \textbf{$1.14\times 10^{-5}$} \\
$Z_{\mathrm{tok}}$ accessibility $A_2$ & RIT $-$ Nonlinear & 0.247 & [0.182, 0.309] & 1.673 & $5.72\times 10^{-6}$ & \textbf{$1.14\times 10^{-5}$} \\
\addlinespace[4pt]
\end{longtable}
\endgroup

\end{document}